\documentclass[twocolumn]{autart}    

\usepackage{graphicx}          

\usepackage{amsmath} 
\usepackage{amssymb}  
\usepackage{pbox} 

\usepackage{tikz} 
\usepackage{pgfplots} 

\usepackage{algpseudocode}
\usepackage{algorithm}
\usepackage{epstopdf}
\usepackage{hyperref} 

\usepackage{color}

\usepackage{mathtools}

\DeclareMathOperator*{\L2}{\mathcal{L}_2}

\begin{document}

\begin{frontmatter}

\title{Determining optimal input-output properties: \\ A data-driven approach \thanksref{footnoteinfo}} 

\thanks[footnoteinfo]{Corresponding author Anne Koch. Tel. +49-711-685-67752. 
Fax +49-711-685-67735.} 

\author{Anne Koch}\ead{anne.koch@ist.uni-stuttgart.de},    
\author{Julian Berberich}\ead{julian.berberich@ist.uni-stuttgart.de},              
\author{Johannes K\"ohler}\ead{johannes.koehler@ist.uni-stuttgart.de},  
\author{Frank Allg\"ower}\ead{frank.allgower@ist.uni-stuttgart.de}

\address{Institute for Systems Theory and Automatic Control, University of Stuttgart, 70569 Stuttgart, Germany.}


\begin{abstract}                          
Due to their relevance in systems analysis and (robust) controller design, we consider the problem of determining control-theoretic system properties of an a priori unknown system from data only. More specifically, we introduce a necessary and sufficient condition for a discrete-time linear time-invariant system to satisfy a given integral quadratic constraint (IQC) over a finite time horizon using only one input-output trajectory of finite length. Furthermore, for certain classes of IQCs, we provide convex optimization problems in form of semidefinite programs (SDPs) to retrieve the optimal, i.e. the tightest, system property description that is satisfied by the unknown system. Finally, we provide bounds on 
the difference between finite and infinite horizon IQCs and illustrate the effectiveness of the proposed scheme in a variety of simulation studies including noisy measurements and a high dimensional system. 
\end{abstract}

\end{frontmatter}

\section{Introduction}

Most established theory on controller design is based on a model of a system. However, with the growing complexity of the plants to be controlled, finding suitable models from first principles quickly becomes an arduous task. Therefore, there has been a rising interest in learning controllers directly from data, commonly referred to as 'data-driven control' approaches. Many such approaches are summarized in \cite{Hou2013}.
While model-based control theory usually provides rigorous mathematical guarantees for the stability and performance of the controlled loop, endowing data-driven methods with the same guarantees is an open field of research. We highlight a few of the very diverse but promising methods and directions in the following (that are not included in \cite{Hou2013}). 

In \cite{Berkenkamp2017} a reinforcement learning algorithm is presented which explicitly considers stability guarantees. In the same direction, the stability of a closed loop represented by a Gaussian Process is examined in \cite{Vinogradska2016}. Both works and references therein are hence examples of works that apply the stability concept from systems and control theory to established (machine) learning approaches. Along the lines of identification for control \cite{Gevers2005}, another approach that recently regained further attention is to identify a model with guaranteed bounds on the uncertainty, which are then accounted for in the robust controller design. A promising work in this direction for linear time-invariant (LTI) systems was presented in \cite{Dean18,Umenberger2019}. However, there are still many open questions aiming at presenting end-to-end guarantees in an automatic data-based control design procedure. 

Another area naturally suited for data-driven concepts is the behavioral context. In \cite{Willems05}, for example, it was shown how a single data trajectory can be used to parametrize all possible future system trajectories for LTI systems. On the basis of this result, data-driven feedback controllers with stability and performance guarantees were introduced in \cite{Persis2019}, with robust controller design presented in \cite{Berberich2019c}. Furthermore, a data-driven model predictive controller was presented in \cite{Coulson2019} with stability guarantees provided in \cite{Berberich2019b}, and determining control-theoretic system properties from data was approached in \cite{Maupong2017,Romer2019a,Yan2019}.
Besides \cite{Maupong2017,Romer2019a,Yan2019}, there have been considerably more approaches to determine system properties such as the $\mathcal{L}_2$-gain (or operator gain), passivity properties, or more generally dissipation inequalities. Why is obtaining control-theoretic properties from data so attractive? Knowledge of certain system properties such as the operator gain or passivity properties allow for the direct application of well-known feedback theorems. Therefore, learning such system properties from data can retain many of the desired advantages of data-driven approaches (simple to apply, no expert knowledge required), while still providing insights into the unknown system and guarantees for the closed-loop behavior with no restrictions on the controller structure. Besides the obvious application of controller design via well-known feedback theorems 
(e.g. \cite{Desoer1975,Megretski1997,Schaft2000,Zames1966}), there have been more specific applications of data-driven system properties for controller validation \cite{Heusden2009}, fault detection and mitigation \cite{Zakeri2019}, and model-free cooperative controller design \cite{Sharf2019}, to highlight some examples.

Due to those reasons, learning system properties from data has recently gained quite some attention. While the estimation of the operator gain has a longer history \cite{Mueller2019,Oomen2014,Poolla1994,Rallo2016,Rojas2012,Wahlberg2011,Wahlberg2010}, more recent approaches for learning system properties extended their consideration to passivity properties \cite{Romer2019c} or more general dissipation inequalities \cite{Maupong2017,Montenbruck2016,Romer2019a,Romer2017a}. All these methods aim at avoiding 
the computational load of identifying a full model and skipping an unnatural fit to a parametric system possibly introducing additional error.
Very generally, nonparametric methods to learn system properties directly from data can be divided into online methods, where iterative experiments on the unknown system are performed, and offline methods, where one has access to previously measured trajectories of the system. While online methods for determining system properties of LTI systems \cite{Mueller2019,Oomen2014,Rallo2016,Rojas2012,Romer2019c,Wahlberg2011,Wahlberg2010} come with their own advantages, assuming that arbitrary iterative experiments can be performed on the system is quite restrictive and possibly time-consuming. Therefore, we consider here an offline approach, which requires only one input-output trajectory, where the input is persistently exciting of suitable order. 

In fact, our approach in this paper is much in line with \cite{Maupong2017}, which introduced the idea of determining dissipativity from a one-shot trajectory on the basis of the behavioral framework. However, their approach resulted in a non-convex indefinite quadratic program which generally is very hard to solve. In \cite{Romer2019a}, verification of dissipativity was boiled down to checking positive semidefiniteness of a single matrix, which thus could be efficiently verified. Similar results were obtained in \cite{Yan2019} in the behavioral framework. In this paper, we provide necessary and sufficient conditions for a discrete-time LTI system to satisfy an integral quadratic constraint (IQC) based only on a definiteness condition on a single data-dependent matrix in Sec.~\ref{sec:main}. IQCs can be used to receive a more informative and tighter description of the unknown system compared to dissipativity properties. This then allows, e.g., for less conservative robust controller design. Therefore, we additionally characterize optimal, i.e., tight system properties and provide SDPs to find these IQCs in Sec.~\ref{sec:iterative}. 
While most of the paper considers the IQC property over a finite time horizon, we infer bounds on the respective property over the infinite time horizon in Sec.~\ref{sec:infinite} and conclude this paper with simulation studies demonstrating the potential of this approach in Sec.~\ref{sec:example}.

\section{Problem Formulation}

\subsection{Notation}
We write $I_n$ for the $n\times n$ identity matrix and $0_{n\times m}$ for an $n\times m$ zero matrix.
If the dimensions of the matrices are clear from the context, we will omit the coefficients $n,m$.
For some matrix $A$, we denote by $A^\perp$ the orthogonal complement of $A$, i.e., a matrix which contains the column vectors spanning the kernel of $A$. The symbol $\otimes$ is used to represent the Kronecker product.
Given a finite sequence $\left\{x_k\right\}_{k=0}^{N-1}$, we define the corresponding Hankel matrix
\begin{align*}
H_L&(x)\coloneqq\begin{pmatrix}x_0 & x_1 & \dots & x_{N-L}\\
x_1 & x_2 & \dots & x_{N-L+1}\\
\vdots & \vdots & \ddots & \vdots\\
x_{L-1} & x_L & \dots & x_{N-1}
\end{pmatrix}.
\end{align*}
We will use $x$ to denote either the sequence itself or the stacked vector containing its components. The Hilbert space $l_2^n$ consists of square summable vector-valued sequences $x=(x_0, x_1, \dots )$, where $\langle x, x \rangle_{l_2^n} = \| x \|_{l_2^n}^2 := \sum_{k=0}^\infty x_k^\top  x_k < \infty$. 
Furthermore, the space of matrix-valued real-rational functions essentially bounded on the unit circle is denoted by $\mathcal{RL}_\infty$. 
For any transfer function $\Psi \in \mathcal{RL}_\infty^{n_r \times (m+p)}$, we let $T_L(\Psi)$ denote the block Toeplitz matrix representing the input-output map of length $L$
\begin{align*}
T_L({\Psi}) = \begin{pmatrix}
g_0^{\Psi} & 0 & 0 & \dots & 0 \\
g_1^{\Psi} & g_0^{\Psi} & 0 & \dots & 0 \\
\vdots & & \ddots & & \vdots \\
g_{L-1}^{\Psi} & g_{L-2}^{\Psi} & \dots & & g_0^{\Psi}%
\end{pmatrix},
\end{align*}
where $\{(g_k^{\Psi})\}_{k=0,1,2,\dots}$ is the impulse response of $\Psi$ with $g_k^{\Psi} \in \mathbb{R}^{n_r \times (m+p)}$, $k=1,2,\dots$. Moreover, $\star$ denotes the convolution operator, e.g. $$r_k = \left(g^{\Psi} \star \begin{pmatrix} u \\ y \end{pmatrix}\right)_k$$ with $u_k \in \mathbb{R}^m$, $y_k \in \mathbb{R}^p$ and $r_k \in \mathbb{R}^{n_r}$.
The adjoint of $X$ is denoted by $X^\star$. 
The subspace $\mathcal{RH}_\infty$ consists of functions in $\mathcal{RL}_\infty$ that are analytic outside the unit circle. The para-hermitian conjugate of a complex matrix-valued function $G(z)$ is $G^\sim(z) := G^\top (z^{-1})$. Note that $G^\sim(e^{j \omega}) =  G^\star(e^{j \omega})$ for all $\omega \in \mathbb{R} \cup \{\pm \infty\}$. 

\subsection{Setup}
We consider discrete-time multiple-input multiple-output (MIMO) LTI systems of order $n$ with $m$ inputs and $p$ outputs. We are interested in the case where no model of the given system is known, but an input-output trajectory of the system is available. Note that we only consider trajectories that come from a minimal realization of an MIMO LTI system. On the basis of that input-output trajectory, we aim to determine input-output properties of the unknown system.
\begin{defn}
\label{def:trajectory}
We say that an input-output sequence $\{(u_k,y_k)\}_{k=0}^{N-1}$ is a trajectory of an LTI system $G$, if there exists an initial condition $x_0\in\mathbb{R}^n$ such that
\begin{align*}
x_{k+1}&=Ax_k+Bu_k,\>\>x(0)=x_0,\\
y_k&=Cx_k+Du_k,
\end{align*}
for $k=0,\dots,N-1$ with $(A,B,C,D)$ defining a minimal realization of $G$.
\end{defn}
To have sufficient information in the initially measured input-output trajectory of our system, we will need the condition that the input of the measured trajectory is persistently exciting in the following sense. 
\begin{defn}\label{def:pe}
We say that a signal $\left\{u_k\right\}_{k=0}^{N-1}$ with $u_k\in\mathbb{R}^m$ is persistently exciting of order $L$, if $\text{rank}\left(H_L(u)\right)\geq mL$.
\end{defn}

Note that Definition~\ref{def:pe} implies $N\geq(m+1)L-1$, yielding a lower bound on the length of the required input-output trajectory for being persistently exciting of order $L$, which also dependents on the number of inputs $m$. Our approach is based on results originally developed in the context of behavioral systems theory in~\cite{Willems05}, which provide a characterization of all trajectories of
an unknown LTI system on the basis of a single input-output trajectory. This result was reformulated in~\cite{Berberich2019a} as follows, providing a simple equivalent characterization of Definition~\ref{def:trajectory} from data. 
\begin{thm}[\cite{Berberich2019a}]\label{thm:Hankel_spans_system}
Suppose $\{u_k,y_k\}_{k=0}^{N-1}$ is a trajectory of an LTI system $G$, where $u$ is persistently exciting of order $L+n$.
Then, $\{ \bar{u}_k,\bar{y}_k \}_{k=0}^{L-1}$ is a trajectory of $G$ if and only if there exists $\alpha\in\mathbb{R}^{N-L+1}$ such that
\begin{align}\label{eq:thm_hankel}
\begin{pmatrix}H_L(u)\\H_L(y)\end{pmatrix}\alpha
=\begin{pmatrix}\bar{u}\\\bar{y}\end{pmatrix}.
\end{align}
\end{thm}
This theorem basically describes that for all linear time-invariant systems any input-output trajectory of the system can be constructed from time-shifts and linear combinations of one measured trajectory of the same system. The only requirement is that the input signal of the measured trajectory is persistently exciting of sufficient order, i.e., entails sufficient information. Therefore, this result constitutes a natural basis for data-based inference of input-output properties.

\subsection{Data-Driven Integral Quadratic Constraints}

We follow the introduction and notation on discrete-time IQCs in \cite{Fry2017,Hu2016,Michalowsky2019}. Generally, Let $P$ be a linear, bounded, self-adjoint operator. Then our system is said to satisfy the IQC defined by the multiplier $P$ if
\begin{align}
\left\langle \begin{pmatrix} u \\ y \end{pmatrix}, P \begin{pmatrix} u \\ y \end{pmatrix} \right\rangle_{l_2} \geq 0 \quad \forall u \in l_2^m.
\label{eq:iqc1}
\end{align}
As introduced in \cite{Megretski1997}, we will in the following characterize $P$ as an LTI system $P \in \mathcal{RL}_\infty^{(m + p) \times (m + p)}$, which includes most multipliers in the literature.

The inner product \eqref{eq:iqc1} can be evaluated in the frequency domain by the Plancheral theorem, which yields 
\begin{align}
\frac{1}{2 \pi} \int_{-\pi}^\pi \begin{pmatrix} \hat{u}(e^{i \omega}) \\ \hat{y}(e^{i \omega}) \end{pmatrix}^\star P(e^{i \omega})  \begin{pmatrix} \hat{u}(e^{i \omega}) \\ \hat{y}(e^{i \omega}) \end{pmatrix}  \mbox{d} \omega,
\label{eq:iqc_frequency}
\end{align}
where $\hat{u}$ and $\hat{y}$ are the discrete-time Fourier transforms of $u$ and $y$, respectively. 
The IQC \eqref{eq:iqc_frequency} can also be evaluated in the time domain based on a factorization $P(z) = \Psi^{\sim}(z) M \Psi(z)$ where $M=M^\top \in \mathbb{R}^{n_r \times n_r}$ and $\Psi \in \mathcal{RH}_\infty^{n_r \times (m+p)}$. 
A factorization of such a form $P(z) = \Psi^{\sim}(z) M \Psi(z)$ is always possible, although it is not unique. 
In fact, one can always construct a factorization from any $P$ such that $\Psi$ is causal and stable \cite{Hu2017}. Hence, throughout the paper, we consider $\Psi$ to be a stable and causal LTI system with zero initial condition, if not otherwise explicitly stated.
This leads us finally to the following definition of IQCs in the discrete time domain \cite{Fry2017}. 
\begin{defn}\label{def:iqc}
We say that an LTI system $G$ satisfies an IQC for a given multiplier $P(z) = \Psi^{\sim}(z) M \Psi(z)$ if 
\begin{align}
\begin{split}
\sum_{k=0}^{h} & r_k^\top M r_k \geq 0 \quad \forall h \geq 0 \\
& \text{with}\quad 
r_k = \left( g^{\Psi} \star \begin{pmatrix} u \\ y \end{pmatrix} \right)_k
\end{split}
\label{eq:iqc}
\end{align}
for all trajectories $\{u_k,y_k\}_{k=0}^{\infty}$ of $G$ with $u \in l^m_2$  
and initial condition $x_0=0$, where $x$ is the state of an arbitrary minimal realization of $G$. 
\end{defn}
Intuitively, computing the signal $r$ corresponds to filtering the input and output signals $\{u_k, y_k\}_{k=0}^\infty$ through an LTI system $\Psi$ with zero initial condition (here simply denoted by the convolution operator with the impulse response of $\Psi$). The time domain IQC is then an inequality ('dissipativity condition') on the filtered output $r$. 

Since we can only measure input-output trajectories over a finite horizon, we introduce the relaxed version of finite-time IQC, or in short $L$-IQC, in this paper. Results for the infinite horizon can be found in Sec.~\ref{sec:infinite}, where the connection of IQCs and $L$-IQCs is discussed. The definition of $L$-IQCs is based on the definition of $L$-dissipativity as introduced in \cite{Maupong2017}. L-dissipativity can be seen as a special subclass of $L$-IQCs, as explained below.  

\begin{defn}\label{def:L_IQC}
We say that an LTI system $G$ satisfies an $L$-IQC for a given $P(z) = \Psi^{\sim}(z) M \Psi(z)$ if 
\begin{subequations}\label{eq:L_IQC}
\begin{align}
\sum_{k=0}^{h} & r_k^\top M r_k \geq 0 \quad \forall h =0, \dots, L-1 
\label{eq:L_IQC_1}\\
& \text{with} \quad 
r_k = \left( g^{\Psi} \star \begin{pmatrix} u \\ y \end{pmatrix} \right)_k
\label{eq:filter}
\end{align}
\end{subequations}
for all trajectories $\{u_k,y_k\}_{k=0}^{L-1}$ of $G$ with initial condition $x_0=0$, where $x$ is the state of an arbitrary minimal realization of $G$. 
\end{defn}

It is actually sufficient for an $L$-IQC to verify that \eqref{eq:L_IQC} holds for the horizon $h = L-1$. 
\begin{prop}\label{prop:iqc_shorter_hor}
An LTI system $G$ satisfies an L-IQC for a given $P(z) = \Psi^{\sim}(z) M \Psi(z)$ if and only if
\begin{align}
\begin{split}
\sum_{k=0}^{L-1} r_k^\top M r_k \geq 0, 
\quad \text{with} \; 
r_k = \left( g^{\Psi} \star \begin{pmatrix} u \\ y \end{pmatrix} \right)_k
\end{split}
\label{eq:L_IQC_L1}
\end{align}
holds for all trajectories $\{u_k,y_k\}_{k=0}^{L-1}$ of $G$ with initial condition $x_0=0$, where $x$ is the state of an arbitrary minimal realization of $G$.
\end{prop}
The proof follows the arguments in the proof of \cite[Prop.~6]{Romer2019a} and is hence omitted here.

As already hinted towards before, dissipativity properties constitute one very important subclass of IQCs, where $\Psi$ and hence $P$ are constant matrices (e.g. $\Psi = I_{m+p}$, $P=M \in \mathbb{R}^{(m+p) \times (m+p)}$). This subclass includes important input-output properties such as the operator gain ($H_\infty$-norm), passivity properties (input- and output-strict passivity), and conic relations. Therefore, any results on IQCs naturally include dissipativity properties. 

Having introduced the notation, problem setup together with some pre-analysis results in this section, in the remainder of the paper we will 
\begin{itemize}
\item show how to verify \eqref{eq:L_IQC} given one input-output trajectory in Sec.~\ref{sec:main},
\item discuss how these methods can be used to find an 'optimal' IQC in Sec.~\ref{sec:iterative},
\item provide bounds on the respective system properties over the infinite time horizon in Sec.~\ref{sec:infinite}, 
\item and prove the practicality of the introduced method on a high dimensional numerical example in Sec.~\ref{sec:example}.
\end{itemize}

\section{Data-Based Characterization of System Properties}
\label{sec:main}

In this section, we provide necessary and sufficient data-based conditions for an LTI system satisfying an $L$-IQC. 
The underlying idea is to replace all possible trajectories of an LTI system 
in the IQC condition~\eqref{eq:L_IQC_L1} by a Hankel matrix containing the measured data making use of Thm.~\ref{thm:Hankel_spans_system}. 
For this purpose, we first define $\{w_k\}_{k=0}^{N-1}$ to be 
a stacked input-output trajectory $\{u_k,y_k\}_{k=0}^{N-1}$ as defined by 
$$w_k = \begin{pmatrix} u_k \\ y_k \end{pmatrix},\; k=0,1,\dots,N-1.$$
Since~\eqref{eq:L_IQC_L1} only needs to hold for trajectories with $x_0=0$, 
we consider trajectories satisfying $w_0=\dots=w_{\nu-1}=0$ for some integer $\nu$. The restriction to this subspace can be equivalently formulated by $\tilde{V}^\nu w=0$ with 
\begin{align*}
\tilde{V}^\nu {=} 
\begin{pmatrix}
I_{(m+p)\nu} & 0_{(m{+}p)\nu\times (m{+}p)(L{-}\nu)}
\end{pmatrix} \in\mathbb{R}^{\nu(m{+}p)\times L(m{+}p)}.
\end{align*}
Further, we define $V^\nu_L(w)=\left(\tilde{V}^\nu H_L(w)\right)^\perp$ to capture this condition via Finsler's lemma in the following theorem, 
which provides necessary and sufficient conditions for an $L$-IQC from data.

\begin{thm}\label{thm:data_driven_IQC}
Suppose $\left\{u_k,y_k\right\}_{k=0}^{N-1}$ is a trajectory of an LTI system $G$. 
\textbf{(i)} 
If $u$ is persistently exciting of order $L+n$ and
\begin{align}\label{eq:data_IQC_LMI}
V^{\nu \top}_L(w) H_L^\top(w) T_L^\top(\Psi) (I_L{\otimes} M) T_L(\Psi) H_L(w)V^\nu_L(w)\succeq0
\end{align}
for some $\nu<L$, then $G$ satisfies the $(L-\nu)$-IQC defined by $P(z) = \Psi^{\sim}(z) M \Psi(z)$. 
\textbf{(ii)}  
If $G$ satisfies the $(L-\nu)$-IQC defined by $P(z) = \Psi^{\sim}(z) M \Psi(z)$, 
then~\eqref{eq:data_IQC_LMI} holds for any $\nu$ with $n\leq\nu< L$.
\end{thm}
\begin{pf}
\textbf{(i)} By applying Finsler's lemma, inequality~\eqref{eq:data_IQC_LMI} is equivalent to
\begin{align}
\alpha^\top H_L^\top(w) T_L^\top(\Psi) (I_L{\otimes} M) T_L(\Psi) H_L(w) \alpha \geq 0
\label{eq:iqc_pf3}
\end{align}
for all $\alpha \in \mathbb{R}^{N-L+1}$ that satisfy $\tilde{V}^\nu H_L(w) \alpha = 0$. Thm.~\ref{thm:Hankel_spans_system} together with the fact that the rows of the linear equations \eqref{eq:thm_hankel} can be arbitrarily permutated, this yields
\begin{align}
\bar{w}^\top T_L^\top(\Psi) (I_L\otimes M) T_L(\Psi) \bar{w} \geq 0
\label{eq:pf_LIQC}
\end{align}
for all trajectories $\{\bar{w}\}_{k=0}^{L-1}$ of G with $\bar{w}_0 = \dots = \bar{w}_{\nu-1} = 0$. This in turn implies 
\begin{align}
\bar{w}^{\prime \top} T_{L-\nu}^\top(\Psi) (I_{L-\nu}\otimes M) T_{L-\nu}(\Psi) \bar{w}^\prime \geq 0
\label{eq:iqc_pf2}
\end{align}
for all trajectories $\{\bar{w}^\prime\}_{k=0}^{L-\nu-1}$ of $G$ with zero initial condition. 

As $\Psi$ is an LTI system with zero initial condition, the output to any given input $\{\bar{w}^\prime\}_{k=0}^{L-\nu-1}$ to $\Psi$ can be cast in matrix notation as $r = T_L(\Psi) \bar{w}^\prime$, which is hence equivalent to the convolution in \eqref{eq:filter}. Therefore, we obtain
\begin{align}\sum_{k=0}^{L-\nu-1} r_k^\top M r_k \geq 0 \quad \text{with} \; 
r_k = \left( g^{\Psi} \star \bar{w}^\prime \right)_k 
\label{eq:iqc_pf}
\end{align}
which, together with the result of Prop.~\ref{prop:iqc_shorter_hor}, results in $G$ satisfying the $(L-\nu)$-IQC as defined in Def.~\ref{def:L_IQC}. 

\textbf{(ii)} Satisfying an $(L-\nu)$-IQC (with $L> \nu$) implies that \eqref{eq:iqc_pf} and hence \eqref{eq:iqc_pf2} holds for all $\{\bar{w}^\prime\}_{k=0}^{L-\nu-1}$ of $G$ with zero initial condition. For any trajectory $\{\bar{w}_k \}_{k=0}^{L-1}$ of $G$, $\tilde{V}^\nu \bar{w} = 0$ together with $n \leq \nu$ imply $x_{\nu} = 0$ where $x$ is the state of an arbitrary minimal realization of $G$. Therefore, \eqref{eq:iqc_pf2} for all $\{\bar{w}^\prime\}_{k=0}^{L-\nu-1}$ of $G$ with zero initial condition implies that \eqref{eq:pf_LIQC} holds for all trajectories $\{\bar{w}\}_{k=0}^{L-1}$ of G with $\bar{w}_0 = \dots = \bar{w}_{\nu-1} = 0$ whenever $n \leq \nu$. Since $H_L(w) \alpha$ with $\alpha \in \mathbb{R}^{N_L+1}$ such that $\tilde{V}^\nu H_L(w) \alpha = 0$ is a subset of all trajectories $\{\bar{w}\}_{k=0}^{L-1}$ of G with $\bar{w}_0 = \dots = \bar{w}_{\nu-1} = 0$ (with equality if $u$ is persistently exciting of order $L+n$, cf.\ \cite{Berberich2019a}), inequality \eqref{eq:iqc_pf3} follows from \eqref{eq:pf_LIQC}.
Applying Finsler's lemma similar to part (i) (as Finsler's lemma provides necessary and sufficient conditions), this in turn yields \eqref{eq:data_IQC_LMI}.   \hfill \qed
\end{pf}

The above theorem can be seen as an extension of \cite[Thm.~7]{Romer2019a} and provides a data-based characterization of an $L$-IQC (Def.~\ref{def:iqc}). While Def.~\ref{def:iqc} requires that \textit{all} possible input-output trajectories satisfy the inequality~\eqref{eq:iqc}, the condition in Thm.~\ref{thm:data_driven_IQC} is based on only \textit{one} measured input-output trajectory of the system. 
Checking an $L$-IQC defined by a multiplier $P$ finally boils down to simply checking a semi-definiteness condition of one matrix~\eqref{eq:data_IQC_LMI}, which can be obtained from data.

There are only two requirements for verifying an $L$-IQC via Thm.~\ref{thm:data_driven_IQC}. More specifically, the sufficient condition in part (i) of Thm.~\ref{thm:data_driven_IQC} requires persistence of excitation of the input to ensure that the Hankel matrix spans the \textit{full} system behavior. 
However, even if the image of $H_L(w)$ does not span the full system behavior, any vector in the image of $H_L(w)$ is still a trajectory of the underlying LTI system. 
Hence, even if the input signal of the available data pair $\{u_k,y_k\}_{k=0}^{L-1}$ is not persistently exciting, we can still infer via part (ii) of Thm.~\ref{thm:data_driven_IQC} that a system does \textit{not} satisfy a specific $L$-IQC.

The other requirement in Thm.~\ref{thm:data_driven_IQC} is an upper bound on the system order $n$ denoted by $\nu$. More specifically, this upper bound $\nu$ is required for the necessary condition in part (ii) of Thm.~\ref{thm:data_driven_IQC}. This bound is used in the matrix $V_L^{\nu}$, which is used to restrict the IQC condition to trajectories with zero initial conditions. Def.~\ref{def:iqc} thus directly explains the requirement of $\nu \geq n$ in part (ii) of Thm.~\ref{thm:data_driven_IQC}. To be more precise, it would actually be sufficient to choose $\nu$ greater or equal to the lag defined by the smallest integer $l \in \mathbb{Z}_{>0}$ such that the observability matrix of $G$ given by $\mathcal{O}_l (A,C) := 
\text{col} ( C , CA , \dots , CA^{l-1})$ 
has rank $n$, which implies $n \geq l$, as also explained in \cite{Coulson2019}. 
While a suitable upper bound is not an assumption for part (i) of Thm.~\ref{thm:data_driven_IQC}, choosing $\nu$ lower than $l$ might result in a violation of~\eqref{eq:data_IQC_LMI} even though the system in fact satisfies the given IQC. For practical applications, $\nu$ can simply be chosen relatively large with the only drawback that the horizon over which an $L$-IQC is guaranteed, i.e. $L-\nu$, decreases. 

We also want to stress here, how particularly simple the condition~\eqref{eq:data_IQC_LMI} for an $L$-IQC is. Not only can the semi-definiteness condition be simply verified by computing, for example, the smallest eigenvalue of the resulting matrix via Matlab functions such as \textit{eigs}, it is also easy to grasp. While $T_L^\top(\Psi) (I_L \otimes M) T_L(\Psi)$ represent the IQC, the Hankel matrix $H_L(w)$ spans the system behavior and $V^\nu_L(w)$ relaxes the conditions to only trajectories with zero initial conditions.

\begin{rem}
Throughout the paper, we assume $\Psi$ to be causal and stable for simplicity, as there always exists a factorization of $P$ such that this is satisfied (\cite[Lemma 1]{Hu2017}). However, Thm.~\ref{thm:data_driven_IQC} can also be extended to acausal multipliers $\Psi$ as Toeplitz matrices can represent causal and acausal LTI operators. Special attention must then be given to the choice of $\nu$. For more details on handling acausal multipliers by Toeplitz matrices, the reader is referred to \cite{Michalowsky2019}.
\end{rem}

\begin{rem}
While Thm.~\ref{thm:data_driven_IQC} includes the multiplier $\Psi$ as a matrix $T_L(\Psi)$ multiplied to $I_L \otimes M$, one could also 
filter the measured trajectory $\{u_k,y_k\}_{k=0}^{L-1}$ by $\Psi$ and apply the results from \cite{Romer2019a} to the filtered signal $\{r_k\}_{k=0}^{L-1}$. However, one would need to account for the order of the filter in the estimate of $\nu$. More importantly, this would not allow to optimize over the filter $\Psi$, which will be done in Sec.~\ref{sec:iterative}. 
\end{rem}

\begin{rem}
\label{sec:noise}
In practice the output measurement is often corrupted by measurement noise, which can be modeled as an instance of a stochastic process $\left\{\varepsilon_k\right\}_{k=0}^{N-1}$.
In this case, a relaxation of the proposed IQC verification analogue to \cite{Romer2019a} can be applied. 
The idea is, on a high level, to approximate the perturbation that the noise causes on the matrix condition from noisy data $\{ \tilde{w}_k \}_{k=0}^{N-1}$, and relax the semi-definiteness condition \eqref{eq:data_IQC_LMI} accordingly by
\begin{align}\label{eq:iqc_noise_relaxation}
V^{\nu\top}_L(\tilde{w}) H_L^\top(\tilde{w}) T_L^\top(\Psi) (I_L {\otimes} M) T_L(\Psi) H_L(\tilde{w})V^\nu_L(\tilde{w})\succeq\delta I
\end{align}
for some $\delta < 0$.
To approximate the influence of the measurement noise, we sample $K$ arbitrary noise instances $\{\varepsilon_k^{(i)}\}_{k=0}^{N-1}$, $i=1, \dots, K$ from the assumed noise distribution offline and compute $\delta$ by 
\begin{align}\label{eq:noise_delta}
\delta=\frac{1}{K}\sum_{i=1}^K\lambda_{\min}\left(V_L^{\nu\top}(\tilde{w}) E_L(\tilde{w},\varepsilon^{(i)})V_L^\nu(\tilde{w})\right)
\end{align}
with 
$D_L(\tilde{w})= H_L^\top(\tilde{w}) T_L^\top(\Psi) (I_L \otimes M) T_L(\Psi) H_L(\tilde{w})$, and $
E_L(\tilde{v},\varepsilon^{(i)}) = D_L(\tilde{\tilde{w}}^{(i)}) - D_L(\tilde{w})$ with $\tilde{\tilde{w}}^{(i)}_k = \tilde{w}_k + ( 0 \; \; \varepsilon_k^{(i)\top} )^\top$.
Summarizing, this leads to the following algorithm.
\begin{algorithm}
\begin{alg}\label{alg:noise}
\normalfont{\textbf{IQCs from noisy measurements}}
\begin{enumerate}
\item Measure data $\{u_k,\tilde{y}_k\}_{k=0}^{N-1}.$
\item Draw $K$ noise samples $\{\varepsilon_k^{(i)}\}_{k=0}^{N-1}$, $i=1,\dots,K$ offline from the noise distribution.
\item Compute $\delta$ using \eqref{eq:noise_delta}.
\item Use~\eqref{eq:iqc_noise_relaxation} for checking the $L$-IQC.
\end{enumerate}
\end{alg}
\end{algorithm}

Noise can also be reduced by e.g. 
averaging (or to taking any convex combination) over signals if more than one trajectory is available. While the above approach cannot provide guarantees on the resulting IQC, it provides very promising results in practice.
\end{rem}

\begin{exmp}
\label{ex:extended_l2}
Let us demonstrate the potential of Thm.~\ref{thm:data_driven_IQC} with a numerical example.
We choose three random $2 \times 2$ MIMO system with system order $n=3$ via the Matlab function \textit{drss(3,2,2)} with the seed \textit{rng(i)}, $i=2,3,4$. 
We choose $\nu = 3$, $L=200$ and $N=500$
together with the IQC from \cite[p.~3147]{Fry2017}:
\begin{align*}
\Psi(z) &= \begin{pmatrix}
B(z) \otimes I_{m} & 0 \\
0 & B(z) \otimes I_{p}
\end{pmatrix} ,\\
M &= \begin{pmatrix}
\gamma^2 X \otimes I_{m} & 0 \\
0 & {-} X \otimes I_{p}
\end{pmatrix},
\end{align*}
where $B(z) = \begin{pmatrix} 1 & \frac{1}{z- \lambda} & \dots & \frac{1}{z- \lambda}^{b-1} \end{pmatrix}^\top$ with $b=3$, $\lambda = 0.5$ and $X = I_3$. We compute the smallest $\gamma$ via bisection such that our unknown systems still satisfy the $L$-IQC. 
We then consider measurements subject to uniform multiplicative noise of the form $\tilde{y_k} = (1+ \varepsilon_k) y_k$ with $\varepsilon_k \in [-\bar{\varepsilon}, \bar{\varepsilon}]$ where $\bar{\varepsilon} > 0$ represents the signal to noise ration (SNR). Fig.~\ref{fig:result_Ex1} presents the results corresponding to an increasing noise level with $K=10$.
\begin{figure}[t]
\begin{center}
\begin{tikzpicture}[scale=0.8]
 \newcommand*{\ax}{0.5}
 \newcommand*{\ay}{.75}
 \newcommand*{\dx}{-0.2}
 \draw[->] (0,0) -- (5,0) node[anchor=west] {};
 \draw[->] (0,0) -- (0,4.25) node[anchor=south,xshift=48,yshift=-15]{};  
\draw (0,\ay*1) -- (-.1,\ay*1) node[anchor=east,yshift=0] {$1$};
\draw (0,\ay*2) -- (-.1,\ay*2) node[anchor=east,yshift=0] {$2$};
\draw (0,\ay*3) -- (-.1,\ay*3) node[anchor=east,yshift=0] {$3$};
\draw (0,\ay*4) -- (-.1,\ay*4) node[anchor=east,yshift=0] {$4$};
\draw (0,\ay*5) -- (-.1,\ay*5) node[anchor=east,yshift=0] {$5$};
\node[anchor=north, rotate=90] at (-1.25,\ay*3) {$\gamma$};
\draw (\ax*2+\dx,0) -- (\ax*2+\dx,-0.1) node[anchor=north] {$0$};
\draw (\ax*4+\dx,0) -- (\ax*4+\dx,-0.1) node[anchor=north] {$0.1$};
\draw (\ax*6+\dx,0) -- (\ax*6+\dx,-0.1) node[anchor=north] {$0.2$};
\draw (\ax*8+\dx,0) -- (\ax*8+\dx,-0.1) node[anchor=north] {$0.3$};
\node[anchor=north] at (\ax*4+\dx,-.65) {$\bar{\varepsilon}$};
\filldraw[blue] (\ax*2+\dx,\ay*2.2734) circle (.075);
\filldraw[green!80!black] (\ax*2+\dx,\ay*3.1889) circle (.075);
\filldraw[red] (\ax*2+\dx,\ay*4.9141) circle (.075);
\filldraw[blue] (\ax*4+\dx,\ay*2.1142) circle (.075);
\filldraw[green!80!black] (\ax*4+\dx,\ay*3.0599) circle (.075);
\filldraw[red] (\ax*4+\dx,\ay*4.8616) circle (.075);
\filldraw[blue] (\ax*6+\dx,\ay*2.0745) circle (.075);
\filldraw[green!80!black] (\ax*6+\dx,\ay*2.9562) circle (.075);
\filldraw[red] (\ax*6+\dx,\ay*4.8755) circle (.075);
\filldraw[blue] (\ax*8+\dx,\ay*2.0348) circle (.075);
\filldraw[green!80!black] (\ax*8+\dx,\ay*2.8170) circle (.075);
\filldraw[red] (\ax*8+\dx,\ay*4.9043) circle (.075);
\draw[blue!95!black, dotted] (5,\ay*2.2734) -- (-.1,\ay*2.2734) node[pos=.01,anchor=south] {};
\draw[green!75!black, dotted] (5,\ay*3.1889) -- (-.1,\ay*3.1889) node[pos=.01,anchor=south] {};
\draw[red!95!black, dotted] (5,\ay*4.9141) -- (-.1,\ay*4.9141) node[pos=.01,anchor=south] {};
\end{tikzpicture}
\caption{Finding an $L$-IQC for three random $2 \times 2$ MIMO systems (red, green, blue) from one input-output trajectories corrupted by measurement noise of different levels with Algorithm~\ref{alg:noise}. 
}
\label{fig:result_Ex1}
\end{center}
\end{figure}
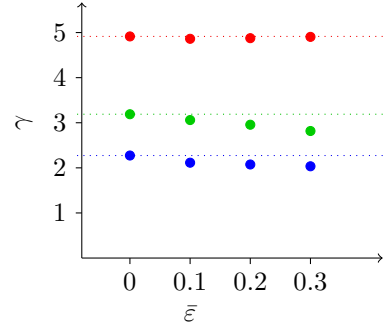
While, very generally, the result deteriorates with larger noise, this small examples supports the claim that the presented approach closely approximates the respective system property even for large noise levels.
\end{exmp}

\section{Data-Driven Inference of Optimal System Properties}
\label{sec:iterative}
In the previous section, we introduced an approach to verify whether an LTI system satisfies an IQC with given $P(z) = \Psi^{\sim}(z) M \Psi(z)$ over the horizon of $(L-\nu)$ with Thm.~\ref{thm:data_driven_IQC}. Usually, however, we are interested in finding some 'optimal' IQC, which within a class of IQCs provides the tightest description of the unknown system. 
For instance, one might want to estimate the $\mathcal{L}_2$-gain of the system, i.e., the minimal $\gamma$ such that the system is dissipative w.r.t.
\begin{align*}
P=\begin{pmatrix}\gamma^2 I_m & 0\\0 & -I_p\end{pmatrix}.
\end{align*}
This can be done via a standard bisection method, or we can state it as a simple SDP reading
\begin{align*}
& \min_{\gamma^2} \gamma^2 \\
& \text{s.t.} \;  
{V_L^\nu}^\top(w) H_L^\top(w) (I_L \otimes P) H_L(w) V_L^\nu(w) \succeq 0.
\end{align*}

Other simple but important system properties are input-strict and output-strict passivity. The excess or shortage of the respective passivity property, i.e., the minimal $\rho$ such that the system is dissipative w.r.t.
\begin{align*}
P=\begin{pmatrix}\rho_{\text{i}} I_m & \frac{1}{2}I_m\\\frac{1}{2}I_m & 0\end{pmatrix}\quad\text{or}\quad
P=\begin{pmatrix}0 & \frac{1}{2}I_m\\\frac{1}{2}I_m & \rho_{\text{o}} I_m\end{pmatrix}
\end{align*}
can again be found via a simple bisection algorithm or an SDP since $P$ is linear in $\rho_i$, $\rho_o$.

The \textit{optimal} IQC within a specific class of parameterized IQCs that a system satisfies in case of the $\mathcal{L}_2$-gain or passivity parameters is quite intuitive. Similarly, we consider a more general optimal IQC within a specified parameterized class of IQCs to be an IQC for which the semidefiniteness condition is tight. For an optimal IQC, there hence exists a non-trivial input-output tuple $\{u_k, y_k\}_{k=0}^{L-1}$ for which \eqref{eq:iqc} holds with equality. The optimal IQC therefore represents a tight description within the given class of IQCs. 

A very important class of IQCs that has been extensively studied in literature are positive-negative (PN) multipliers for which there exists a factorization such that
\begin{align}
M_\gamma = \begin{pmatrix} \gamma^2 I_{n_{r_1}} & 0 \\ 0 & -I_{n_{r_2}} \end{pmatrix}, \> \Psi = \begin{pmatrix} \Psi_{11} & \Psi_{12} \\ \Psi_{21} & \Psi_{22}\end{pmatrix}, 
\label{eq:iqc_class}
\end{align}
with $\Psi_{11} \in \mathcal{RH}_\infty^{n_{r_1} \times m}$, $\Psi_{12} \in \mathcal{RH}_\infty^{n_{r_1} \times p}$, $\Psi_{21}\in \mathcal{RH}_\infty^{n_{r_2} \times m}$ and $\Psi_{22}\in \mathcal{RH}_\infty^{n_{r_2} \times p}$.
Let us further only consider filters $\Psi$ for which $\Psi_{12} = 0$ (cf.\ triangular factorization for positive-negative multipliers as discussed in~\cite{Carrasco2018}), $\Psi_{11}(z)$ fixed, and $\Psi_{21}(z)$, $\Psi_{22}(z)$ linearly parameterized, i.e. 
\begin{align}
\Psi(z) &= \begin{pmatrix} \Psi_{11}(z) & 0 \\ \Psi_{21}(z) & \Psi_{22}(z) \end{pmatrix}, \label{eq:iqc_class2}\\
\Psi_{21}(z) &= \sum_{k=0}^b c^{(21)}_k B^{(21)}_k(z), \;\;
\Psi_{22}(z) = \sum_{k=0}^b c^{(22)}_k B^{(22)}_k(z). \notag
\end{align}
Here, $B^{(21)}_k(z)$, $B^{(22)}_k(z)$, $k=0,\dots,b$ are fixed basis functions and $c^{(21)}_k \in \mathbb{R}^{n_{r_2} \times m}$, $c^{(22)}_k \in \mathbb{R}^{n_{r_2} \times p}$ are free parameters. Popular basis functions include, e.g., $\left( 1, (z+\lambda)^{-1}, (z+\lambda)^{-2}, \dots, (z+\lambda)^{-b} \right)$ with $|\lambda| < 1$ fixed. 
The corresponding Toeplitz matrices of $\Psi_{21}$ and $\Psi_{22}$ can be computed as 
\begin{align*}
T_L(\Psi_{21}) 
&= \sum_{k=0}^b c^{(21)}_k T_L(B^{(21)}_k), \\
T_L(\Psi_{22}) 
&= \sum_{k=0}^b c^{(22)}_k T_L(B^{(22)}_k).
\end{align*}
Note that $T_L(B^{(21)}_k)$, $T_L(B^{(22)}_k)$ are block Toeplitz matrices, possibly non-square. In the following, we present a semi-definite program (SDP) which computes the optimal IQC within the aforementioned class of IQCs that an a priori unknown LTI system satisfies from only one input-output trajectory.

To improve readability of the main results, we rearrange vectors and matrices and denote 
\begin{align*}
\tilde{\tilde{V}}^\nu &= \begin{pmatrix}
I_{m\nu} &  0_{m\nu \times m(L-\nu)} & 0_{m\nu \times p\nu} & 0_{m\nu \times p(L-\nu)} \\
0_{p\nu \times m\nu} &  0_{p\nu \times m(L-\nu)} & I_{p\nu} & 0_{p\nu \times p(L-\nu)}
\end{pmatrix}, \\
H^\nu_u &= H_L(u) \left( \tilde{\tilde{V}}^\nu \begin{pmatrix}
H_L(u) \\ H_L(y)
\end{pmatrix}
\right)^\perp, \\
H^\nu_y &= H_L(y) \left( \tilde{\tilde{V}}^\nu \begin{pmatrix}
H_L(u) \\ H_L(y)
\end{pmatrix}
\right)^\perp,
\end{align*} and $T_{11} = T_L(\Psi_{11})$, $T_{21}(c) = T_L(\Psi_{21}(c^{21}))$, $T_{22}(c) = T_L(\Psi_{22}(c^{22}))$ to emphasize the parameter dependence of $\Psi_{21}$ and $\Psi_{22}$. 

We are now interested in finding the minimum $\gamma^2$ over all parameterized multipliers \eqref{eq:iqc_class} with \eqref{eq:iqc_class2} 
such that the $L$-IQC holds for the unknown LTI system on the basis of only one input-output trajectory. This can be interpreted as finding the tightest (dynamic) cone, i.e. the cone with minimal radius $\gamma$, that our input-output system is confined to.

\begin{thm}\label{thm:optimal}
Suppose $\{u_k, y_k\}_{k=0}^{N-1}$ is a trajectory of an LTI system $G$, $u$ is persistently exciting of order $L+n$ and $n \leq \nu < L$. 
The smallest $\gamma^2$ such that $G$ satisfies the $(L-\nu)$-IQC \eqref{eq:iqc_class}-\eqref{eq:iqc_class2} can be computed by
\begin{align}
& \min_{\gamma^2, c} \gamma^2 \quad \text{s.t.} 
\label{eq:sdp} \\ 
& \begin{pmatrix}
I & (T_{21}(c) H_u^\nu + T_{22}(c) H_y^\nu)^\top \\
T_{21}(c) H_u^\nu + T_{22}(c) H_y^\nu & \gamma^2 H_u^{\nu \top} T_{11}^\top T_{11} H_u^\nu
\end{pmatrix} \succeq 0. \notag
\end{align}
\end{thm}

\begin{pf}
Rearranging the terms in $u$ and $y$ in the result in Thm.~\ref{thm:data_driven_IQC}, the system $G$ satisfies an $(L-\nu)$-IQC defined by \eqref{eq:iqc_class} and \eqref{eq:iqc_class2} if 
\begin{align*}
\begin{pmatrix} H_u^\nu \\ H_y^\nu \end{pmatrix}^{\top}
\hspace{-4pt}\begin{pmatrix} T_{11} & 0 \\ T_{21}(c) & T_{22}(c) \end{pmatrix}^\top \hspace{-4.5pt}{M_\gamma}{\otimes}{I_L} \hspace{-2pt}\begin{pmatrix} T_{11} & 0 \\ T_{21}(c) & T_{22}(c) \end{pmatrix}\hspace{-2pt}
\begin{pmatrix} H_u^\nu \\ H_y^\nu \end{pmatrix} 
\end{align*}
is positive semidefinite. This yields the equivalent semidefiniteness condition
\begin{align*}
& \gamma^2 H_u^{\nu \top} T_{11}^\top T_{11} H_u^\nu\\
&{-} \begin{pmatrix}
T_{21}(c) H_u^\nu {+} T_{22}(c) H_y^\nu
\end{pmatrix}^\top \begin{pmatrix}
T_{21}(c) H_u^\nu {+} T_{22}(c) H_y^\nu
\end{pmatrix}
\succeq 0. 
\end{align*}
Using the Schur complement, we can rewrite this problem as the SDP given in \eqref{eq:sdp}. \hfill \qed
\end{pf}
The resulting optimization problem \eqref{eq:sdp} is hence an SDP with $2(b+1)+1$ decision variables that can be solved efficiently using standard solvers. 

One very important system property that falls into the class of IQCs defined by \eqref{eq:iqc_class} with \eqref{eq:iqc_class2} are conic relations. 
A system $G$ is said to be confined to a conic region characterized by the real constants $C \in \mathbb{R}^{(p \times m)}$ and $\gamma \geq 0$ if 
the system is dissipative with respect to $P = \Psi_c^\top M_\gamma \Psi_c$ with
\begin{align}
\Psi_c 
=\begin{pmatrix}
I & 0 \\ -C & I
\end{pmatrix}.
\label{eq:conic_filter}
\end{align}

While in \cite{Romer2019c} local convergence towards the tightest cone (minimal $\gamma$) could be obtained via iterative experiments, Thm.~\ref{thm:optimal} provides means to compute the tightest cone describing the a priori unknown LTI system from only one input-output trajectory.

\begin{cor}
\label{cor:conic}
Suppose $\{u_k, y_k\}_{k=0}^{N-1}$ is a trajectory of an LTI system G, $u$ is persistently exciting of order $L+n$ and $n \leq \nu < L$. 
The smallest $\gamma^2$ such that $G$ is confined to the cone described by $(C, \gamma^2)$ over the horizon $(L-\nu)$ can be computed by
\begin{align*}
\begin{split}
& \min_{\gamma^2, C} \gamma^2 \quad \text{s.t.} \\ 
& \begin{pmatrix}
I & (-C_L H_u^\nu + H_y^\nu)^\top \\
-C_L H_u^\nu +H_y^\nu & \gamma^2 H_u^{\nu \top} H_u^\nu
\end{pmatrix} \succeq 0
\end{split}
\end{align*}
with $C_L = I_L \otimes C$.
\end{cor}
Very generally, while the $\L2$-gain might be quite large for some systems, a conic description can decrease $\gamma$ significantly and allowing for a dynamic filter provides an even smaller $\gamma$.

\begin{exmp}
We reconsider the three randomly generated $2 \times 2$ MIMO systems 
from Ex.~\ref{ex:extended_l2}. We choose again $\nu = 3$, $L = 200$ and $N=500$ and we determine the tightest $L$-IQC considering the $\L2$-gain and the parameterization \eqref{eq:iqc_class2} with basis functions $\left( 1, (z+\lambda)^{-1}, (z+\lambda)^{-2}, \dots, (z+\lambda)^{-b} \right)$, $\lambda = 0.8$, $b=0,1,2,3$ for $\Psi_{21}$, and $\Psi_{11}(z) = \Psi_{22}(z) = I_2$. Note that $b=0$ represents conicity (cf.\ Cor.~\ref{cor:conic}). The results are illustrated in Fig.~\ref{fig:result_Ex3} and show that allowing for a conic description already significantly improves the radius $\gamma$ compared to computing the $\L2$-gain and hence decreases the conservatism in the respective feedback theorem \cite{Zames1966}. As expected, we can decrease the radius $\gamma$ even more by allowing for additional dynamics in the filter. To show that we indeed found the best transformation described by the parameters $c$, Fig.~\ref{fig:cone} depicts how the minimal $\gamma$ varies locally with perturbation of the optimal parameterization $c^\star_k \pm 2$, $k=1,\dots,8$ for the random system colored in blue and $b=1$.

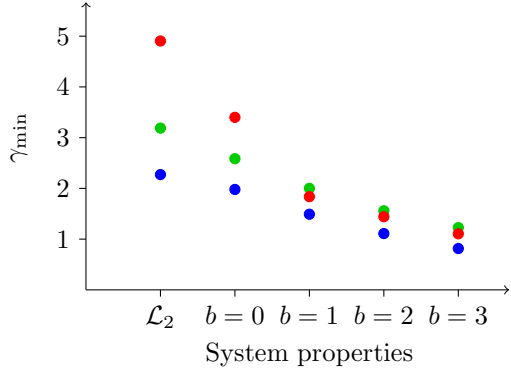
\begin{figure}[t]
\begin{tikzpicture}[scale=0.9]
 \newcommand*{\ax}{0.55}
 \newcommand*{\ay}{.75} 
 \newcommand*{\dx}{0}
 \draw[->] (0,0) -- (6.25,0) node[anchor=west] {};
 \draw[->] (0,0) -- (0,4.25) node[anchor=south,xshift=48,yshift=-15]{};  
\draw (0,\ay*1) -- (-.1,\ay*1) node[anchor=east,yshift=0] {$1$};
\draw (0,\ay*2) -- (-.1,\ay*2) node[anchor=east,yshift=0] {$2$};
\draw (0,\ay*3) -- (-.1,\ay*3) node[anchor=east,yshift=0] {$3$};
\draw (0,\ay*4) -- (-.1,\ay*4) node[anchor=east,yshift=0] {$4$};
\draw (0,\ay*5) -- (-.1,\ay*5) node[anchor=east,yshift=0] {$5$};
\node[anchor=north, rotate=90] at (-1.25,\ay*3) {$\gamma_{\min}$};

\draw (\ax*2+\dx,0) -- (\ax*2+\dx,-0.1) node[anchor=north] {$\L2$};
\draw (\ax*4+\dx,0) -- (\ax*4+\dx,-0.1) node[anchor=north] {$b=0$};
\draw (\ax*6+\dx,0) -- (\ax*6+\dx,-0.1) node[anchor=north] {$b=1$};
\draw (\ax*8+\dx,0) -- (\ax*8+\dx,-0.1) node[anchor=north] {$b=2$};
\draw (\ax*10+\dx,0) -- (\ax*10+\dx,-0.1) node[anchor=north] {$b=3$};
\node[anchor=north] at (\ax*6+\dx,-.65) {System properties};
\filldraw[blue] (\ax*2+\dx,\ay*2.2724) circle (.075);
\filldraw[green!80!black] (\ax*2+\dx,\ay*3.1883) circle (.075);
\filldraw[red] (\ax*2+\dx,\ay*4.9052) circle (.075);
\filldraw[blue] (\ax*4+\dx,\ay*1.9791) circle (.075);
\filldraw[green!80!black] (\ax*4+\dx,\ay*2.5857) circle (.075);
\filldraw[red] (\ax*4+\dx,\ay*3.4005) circle (.075);
\filldraw[blue] (\ax*6+\dx,\ay*1.4913) circle (.075);
\filldraw[green!80!black] (\ax*6+\dx,\ay*2.0013) circle (.075);
\filldraw[red] (\ax*6+\dx,\ay*1.8368) circle (.075);
\filldraw[blue] (\ax*8+\dx,\ay*1.1098) circle (.075);
\filldraw[green!80!black] (\ax*8+\dx,\ay*1.5588) circle (.075);
\filldraw[red] (\ax*8+\dx,\ay*1.4420) circle (.075);
\filldraw[blue] (\ax*10+\dx,\ay*0.8150) circle (.075);
\filldraw[green!80!black] (\ax*10+\dx,\ay*1.2269) circle (.075);
\filldraw[red] (\ax*10+\dx,\ay*1.1045) circle (.075);
\end{tikzpicture}
\caption{Three random systems with $\L2$-gain, tightest cone and best dynamic parameterization.}
\label{fig:result_Ex3}
\end{figure}

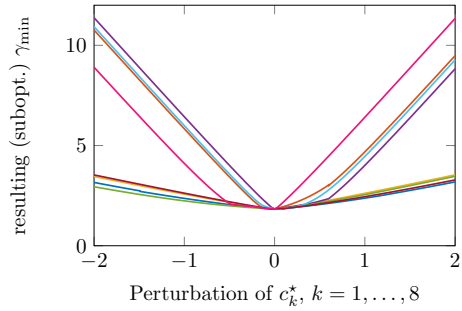
\begin{figure}[]%
\centering
\definecolor{mycolor1}{rgb}{0.00000,0.44700,0.74100}%
\definecolor{mycolor2}{rgb}{0.85000,0.32500,0.09800}%
\definecolor{mycolor3}{rgb}{0.92900,0.69400,0.12500}%
\definecolor{mycolor4}{rgb}{0.49400,0.18400,0.55600}%
\definecolor{mycolor5}{rgb}{0.46600,0.67400,0.18800}%
\definecolor{mycolor6}{rgb}{0.30100,0.74500,0.93300}%
\definecolor{mycolor7}{rgb}{0.63500,0.07800,0.18400}%
\definecolor{mycolor8}{rgb}{0.93500,0.07800,0.48400}%
\begin{tikzpicture}[scale=0.8]

\begin{axis}[%
width=6cm,
height=4cm,
at={(0cm,0cm)},
scale only axis,
xmin=-2,
xmax=2,
ymin=0,
ymax=12,
xlabel={Perturbation of $c_k^\star$, $k=1, \dots, 8$},
ylabel={resulting (subopt.) $\gamma_{\min}$},
axis background/.style={fill=white},
legend style={legend cell align=left, align=left, draw=white!15!black}
]
\addplot [color=mycolor1, thick]
  table[row sep=crcr]{%
-2	3.15741128589865\\
-1.832	3.01514506398904\\
-1.677	2.88683485845091\\
-1.534	2.77136822387275\\
-1.492	2.73804581173232\\
-1.49	2.70942698405588\\
-1.348	2.60519236191453\\
-1.211	2.50744673123295\\
-1.079	2.4160683077324\\
-0.952	2.33090497316176\\
-0.829	2.25113835905831\\
-0.709	2.17601341298486\\
-0.592	2.10543522930741\\
-0.477	2.03872051940492\\
-0.364	1.97580661624218\\
-0.345	1.96548986816894\\
-0.344	1.98463108095053\\
-0.26	1.94542427033119\\
-0.176	1.90878330050469\\
-0.0899999999999999	1.87381644636448\\
0.00300000000000011	1.83839955952843\\
0.0129999999999999	1.83651962830662\\
0.0489999999999999	1.84553525517866\\
0.085	1.85710296270133\\
0.13	1.87416928516826\\
0.188	1.89882290833212\\
0.258	1.93118479893892\\
0.337	1.97029027209496\\
0.422	2.01495798898583\\
0.512	2.0648870250641\\
0.606	2.11969765517591\\
0.704	2.17952498309534\\
0.806	2.24449090808498\\
0.913	2.31537203003822\\
1.025	2.39232056730193\\
1.14	2.47403860055968\\
1.271	2.5761689982605\\
1.413	2.6898240503955\\
1.565	2.81445814416163\\
1.728	2.95110414763357\\
1.904	3.10167440534947\\
2	3.18497540504786\\
};

\addplot [color=mycolor2, thick]
  table[row sep=crcr]{%
-2	10.7606530566846\\
-1.493	8.252908563812\\
-1.134	6.48815445470454\\
-0.871	5.20625060952848\\
-0.673999999999999	4.25691380847783\\
-0.525	3.54963172697536\\
-0.413	3.02852659055232\\
-0.331	2.65717240173347\\
-0.271000000000001	2.3954545400019\\
-0.228	2.21751244026721\\
-0.196	2.09434054166682\\
-0.172000000000001	2.01033295894592\\
-0.151999999999999	1.94827044917879\\
-0.135	1.9028739686367\\
-0.119999999999999	1.86938988023699\\
-0.113	1.85601211660194\\
-0.112	1.87292389285233\\
-0.101000000000001	1.86774931192529\\
-0.0879999999999992	1.85858110421272\\
-0.0779999999999994	1.85206632721422\\
-0.00900000000000034	1.8329729579744\\
-0.00500000000000078	1.83506428707441\\
0.00900000000000034	1.84755039082164\\
0.0329999999999995	1.87183483830796\\
0.0609999999999999	1.9034787714601\\
0.093	1.943234666525\\
0.125999999999999	1.98792335337729\\
0.157	2.03359555216244\\
0.194000000000001	2.09249443862615\\
0.231999999999999	2.15724735411231\\
0.271000000000001	2.22822697253209\\
0.311	2.30587258903235\\
0.351000000000001	2.38856320083126\\
0.391	2.47650133094302\\
0.431000000000001	2.56997772112029\\
0.471	2.66939555926573\\
0.51	2.77255511069158\\
0.548	2.87956999838993\\
0.585000000000001	2.99056690888333\\
0.599	3.03445409366282\\
0.6	3.00779481326109\\
0.647	3.18250731117998\\
0.698	3.38107609340567\\
0.755000000000001	3.61236424897512\\
0.82	3.88602129162516\\
0.894	4.20781556627711\\
0.98	4.59237906976772\\
1.081	5.05488297589395\\
1.202	5.62013283946555\\
1.349	6.31824038172893\\
1.531	7.19416506250399\\
1.761	8.31288727647538\\
2	9.48420881252523\\
};

\addplot [color=mycolor3, thick]
  table[row sep=crcr]{%
-2	3.45625709165804\\
-1.751	3.23022169813587\\
-1.525	3.02809554699974\\
-1.318	2.84598155831693\\
-1.128	2.68182035978338\\
-0.952	2.53273970274986\\
-0.789	2.39762348421704\\
-0.637	2.27454500716453\\
-0.494	2.16165852672419\\
-0.359	2.05797599040951\\
-0.231	1.96253818550528\\
-0.11	1.87513447823245\\
-0.073	1.84899067097647\\
-0.0720000000000001	1.8668645088041\\
-0.044	1.85296690414957\\
-0.0220000000000002	1.84455278100758\\
-0.00300000000000011	1.83978213489207\\
0.00499999999999989	1.83948471994211\\
0.0310000000000001	1.842245368187\\
0.0430000000000001	1.8450285587829\\
0.044	1.82739777904481\\
0.125	1.88558511752711\\
0.214	1.95238977848645\\
0.314	2.03038843182788\\
0.426	2.12072799077068\\
0.552	2.22536605225905\\
0.695	2.34717355920661\\
0.857	2.48824786739447\\
1.041	2.65159097149819\\
1.25	2.84026337788693\\
1.487	3.05736233386824\\
1.756	3.30693719129059\\
2	3.5356200774582\\
};

\addplot [color=mycolor4, thick]
  table[row sep=crcr]{%
-2	11.3764175875367\\
-1.436	8.58129241798722\\
-1.051	6.68417759658434\\
-0.776	5.34001615388794\\
-0.574	4.36348642472992\\
-0.420999999999999	3.63459306162268\\
-0.304	3.08781424842954\\
-0.214	2.67758872823505\\
-0.144	2.36874944220112\\
-0.0899999999999999	2.14043486091068\\
-0.048	1.97239511245152\\
-0.0180000000000007	1.86010790157606\\
-0.0169999999999995	1.8755470743751\\
-0.00500000000000078	1.84907370670038\\
0.00300000000000011	1.8353306135844\\
0.0340000000000007	1.85175420698575\\
0.0850000000000009	1.88113896735398\\
0.138999999999999	1.91492591993363\\
0.192	1.95075460299616\\
0.244	1.98866617287479\\
0.295	2.02870452818237\\
0.323	2.05197073284618\\
0.324	2.03257801113916\\
0.372999999999999	2.08217879121939\\
0.425000000000001	2.1381663926954\\
0.48	2.20083958222558\\
0.538	2.2704433875251\\
0.599	2.34716856704428\\
0.622999999999999	2.43163050565212\\
0.666	2.58998280171346\\
0.714	2.77582483346286\\
0.768000000000001	2.99446302608186\\
0.829000000000001	3.25131614444834\\
0.899000000000001	3.55619906079353\\
0.981	3.923812011026\\
1.079	4.37401235627499\\
1.198	4.93189715793793\\
1.345	5.63254306534719\\
1.53	6.5260069834693\\
1.768	7.68728798327149\\
2	8.82719054185071\\
};

\addplot [color=mycolor5, thick]
  table[row sep=crcr]{%
-2	2.94305286367493\\
-1.862	2.83345178530733\\
-1.733	2.73389624353344\\
-1.612	2.64337013516057\\
-1.498	2.56089001198739\\
-1.389	2.48483053165508\\
-1.285	2.41504149706038\\
-1.186	2.35132608819337\\
-1.09	2.29223852549554\\
-0.997	2.23766305146248\\
-0.906	2.18690198414412\\
-0.816	2.13933779775955\\
-0.727	2.09492100800238\\
-0.639	2.05357125948378\\
-0.551	2.01475587063524\\
-0.462	1.97803792137911\\
-0.373	1.94382830813787\\
-0.285	1.91245281482809\\
-0.2	1.88454804401226\\
-0.124	1.86194060358008\\
-0.0640000000000001	1.84638599841783\\
-0.0190000000000001	1.8369912731578\\
-0.0110000000000001	1.83636911312622\\
0.0209999999999999	1.84602149147817\\
0.085	1.86785031623419\\
0.102	1.87352733510333\\
0.103	1.85552143388321\\
0.187	1.9107195696717\\
0.275	1.97132082858733\\
0.367	2.03747325629465\\
0.465	2.11079295274936\\
0.57	2.19226717684729\\
0.683	2.28291259953013\\
0.805	2.38376093183788\\
0.938	2.49670588596608\\
1.085	2.62459136096554\\
1.248	2.76947905054454\\
1.431	2.93526647541431\\
1.638	3.12595514188414\\
1.874	3.34654270439154\\
2	3.46542874674652\\
};

\addplot [color=mycolor6, thick]
  table[row sep=crcr]{%
-2	10.9299358573427\\
-1.448	8.19480529100765\\
-1.073	6.34756828653924\\
-0.808	5.05306985159952\\
-0.617000000000001	4.13087175027695\\
-0.478	3.47044327365217\\
-0.377000000000001	3.00107877378584\\
-0.304	2.67201620209126\\
-0.25	2.43853708342816\\
-0.209	2.27098733216789\\
-0.177	2.14956423635104\\
-0.151999999999999	2.06304552609901\\
-0.131	1.99789688643345\\
-0.125	1.98077106766059\\
-0.124000000000001	1.998237494084\\
-0.106999999999999	1.96705317999473\\
-0.0869999999999997	1.93532632474267\\
-0.0649999999999995	1.90476521387702\\
-0.0440000000000005	1.879384751936\\
-0.0269999999999992	1.86211044365245\\
-0.00300000000000011	1.84165567883277\\
0.00900000000000034	1.83271898285642\\
0.0589999999999993	1.84786680227699\\
0.134	1.87311257983306\\
0.156000000000001	1.8819041136614\\
0.178000000000001	1.89390322981547\\
0.195	1.90118619186127\\
0.196	1.8831012062383\\
0.221	1.91003537585569\\
0.246	1.94033036309832\\
0.272	1.97553563646351\\
0.298	2.01467467571169\\
0.324	2.05792128825398\\
0.351000000000001	2.10735746203824\\
0.378	2.16156426822744\\
0.406000000000001	2.22294459132064\\
0.435	2.29213443846232\\
0.465	2.36971706096283\\
0.497	2.45908279866459\\
0.529999999999999	2.55812591342377\\
0.566000000000001	2.67368687617223\\
0.603999999999999	2.80350120877652\\
0.646000000000001	2.95537746893707\\
0.692	3.13055191145629\\
0.743	3.33392239303521\\
0.801	3.574831492356\\
0.868	3.86327320887444\\
0.946	4.2096073791878\\
1.038	4.62890759695711\\
1.149	5.14591879835521\\
1.285	5.79074987298723\\
1.455	6.60837625002757\\
1.672	7.66380656210064\\
1.957	9.06192810486113\\
2	9.27367227204464\\
};

\addplot [color=mycolor7, thick]
  table[row sep=crcr]{%
-2	3.53294387682285\\
-1.697	3.25245436043559\\
-1.426	3.00463694990269\\
-1.183	2.78545413461109\\
-0.965	2.59183246708721\\
-0.77	2.42162250131707\\
-0.595	2.27184715471407\\
-0.44	2.14212403848972\\
-0.302	2.0295447538622\\
-0.18	1.93291195175914\\
-0.073	1.85099961600832\\
-0.0350000000000001	1.82268288618283\\
-0.0340000000000003	1.84032828066722\\
-0.016	1.8396141602807\\
0.00800000000000001	1.84044710922411\\
0.0340000000000003	1.84587483650533\\
0.0680000000000001	1.8554526316742\\
0.0880000000000001	1.86201756843002\\
0.089	1.84401969048105\\
0.177	1.89194541996118\\
0.267	1.94359668889159\\
0.36	1.99963891218335\\
0.456	2.06018113806271\\
0.556	2.12597570936018\\
0.66	2.19715317345336\\
0.769	2.27453208544805\\
0.884	2.35899282624517\\
1.006	2.45146762611551\\
1.135	2.5521391301653\\
1.273	2.66276209914332\\
1.421	2.78436637240607\\
1.581	2.91884314531031\\
1.754	3.06728861384462\\
1.942	3.2316696507457\\
2	3.28295128367695\\
};

\addplot [color=mycolor8, thick]
  table[row sep=crcr]{%
-2	8.90565438989262\\
-1.693	7.40276023631139\\
-1.462	6.28283500154002\\
-1.282	5.42106293573284\\
-1.139	4.74720908749763\\
-1.023	4.21114208750558\\
-0.926	3.77332640750244\\
-0.843999999999999	3.41346406860764\\
-0.774000000000001	3.11613871660491\\
-0.712	2.86253572394211\\
-0.657	2.6470722674273\\
-0.608000000000001	2.46417404520905\\
-0.564	2.30843435751125\\
-0.523	2.17153786442063\\
-0.487	2.05858566372589\\
-0.486000000000001	2.07621111518903\\
-0.475	2.06869944483271\\
-0.465999999999999	2.05975207995007\\
-0.458	2.04834496835747\\
-0.454000000000001	2.04323905742072\\
-0.422000000000001	2.02077831634539\\
-0.420999999999999	2.04031905918939\\
-0.306000000000001	1.97931683579479\\
-0.198	1.92681985335458\\
-0.151	1.9047146412024\\
-0.149000000000001	1.88506748752039\\
-0.0229999999999997	1.83947261783709\\
-0.00300000000000011	1.83575191656135\\
0.00200000000000067	1.84270828047608\\
0.0120000000000005	1.85992337273091\\
0.0129999999999999	1.84500144268697\\
0.0500000000000007	1.98798474972916\\
0.0960000000000001	2.17529966634545\\
0.153	2.41730103051586\\
0.224	2.72917859291027\\
0.31	3.11763026069425\\
0.415000000000001	3.60284583155454\\
0.544	4.21020916820095\\
0.704000000000001	4.97502184899029\\
0.904999999999999	5.94749191580528\\
1.163	7.20760513180705\\
1.501	8.87046819713355\\
1.956	11.1210800050333\\
2	11.3392074793172\\
};
\end{axis}
\end{tikzpicture}
\caption{Illustration of optimal parameterizations $c_k^\star$, $k=1, \dots, 8$.}%
\label{fig:cone}
\end{figure}
It is interesting to note that the minimal radius $\gamma_{\min}$ plotted over the different center parameters is continuous but not necessarily differentiable. Further information on this fact can be found in \cite{Romer2019c}.
\end{exmp}

\begin{rem}\label{rem:nonlinearity_measure}
Another special case that falls in the above category of IQCs are dynamic cones, or also called 'nonlinearity measures', as explained for example in \cite{Schweickhardt2007}. To arrive at a dynamic cone description, we choose $\Psi_{11}(z)=I_{m}$ and $\Psi_{22}(z)=I_{p}$, which yields
\begin{align*}
P(z) = \begin{pmatrix} \gamma^2 I - \Psi_{21}^\top(z) \Psi_{21}(z) & \Psi_{21}(z) \\
\Psi_{21}(z) & -I
\end{pmatrix}.
\end{align*}
The smallest radius $\gamma$ over the set of possible LTI approximations is hence a measure on how nonlinear the system at hand is. If the underlying system is linear, the same IQC measures the distance with respect to another linear system. Such a measure can be useful in model reduction of high dimensional systems, for example, and corresponding robust controller design. Hence, for a given reduced-order model $\Psi_{21}(z)$, we can compute the approximation error $\gamma$. Furthermore, by parameterizing $\Psi_{21}(z)$ through basis functions as described above, we identify the closest LTI model within a given class. Hence, Thm.~\ref{thm:optimal} enables what we call here lower-order model approximation (contributing towards 'data-driven model reduction' \cite{Scarciotti2017}). The limitation yet is that we have to parametrize the reduced model linearly to receive convergence guarantees, but with the caveat that we at the same time receive a measure $\gamma$ describing the worst-case deviation between full-order and reduced-order model.
\end{rem}

\begin{exmp}
To underline Rem.~\ref{rem:nonlinearity_measure}, we consider the following $7^{\text{th}}$ order system
\begin{align*}
G(z) = \begin{pmatrix} \frac{2}{z+0.51} & \frac{1}{z+0.19}+\frac{1}{z+0.21} \\ \frac{1}{z+0.55} + \frac{2}{z+0.2} & \frac{2}{z+0.52} + \frac{3}{z+0.5}
\end{pmatrix},
\end{align*}
which has an $\L2$-gain of $11.9$. We
parameterize $\Psi_{21}(z)$ with the basis functions $(1,(z+0.5)^{-1}, (z+0.2)^{-1})$ and choose $\Psi_{11}(z) = \Psi_{22}(z) = I_2$. Choosing $\nu = 10$, $L = 110$ and $N=300$ we retrieve the optimal parameterization (i.e. the minimal $\gamma$) for 
\begin{align*}
\Psi_{12}(z) = 
\begin{pmatrix} 2.1 & 0.0 \\ 1.3 & 5.2 \end{pmatrix} \frac{1}{1+0.5 z}
+ \begin{pmatrix} -0.1 & 2.0 \\ \phantom{-}1.7 & 0.2 \end{pmatrix} \frac{1}{1+0.2 z}
\end{align*}
which closely approximates $G$ with the guaranteed approximation error $\gamma = 0.05$ and model order $4<7$.
\end{exmp}

In this section, we have introduced an SDP for finding the optimal IQC parametrized by NP multipliers \eqref{eq:iqc_class} with \eqref{eq:iqc_class2}, which represents a quite general and important class of IQCs. Another parameterization that is equally applicable is, e.g. (cf. \cite[Eq. (30b)]{Veenman2013}),
\begin{align*}
M = \begin{pmatrix} 0 & I \\ I & - \gamma \end{pmatrix}, \quad \Psi(z) = \begin{pmatrix}
\Psi_{11}(z) & \Psi_{12}(z) \\
0 & \Psi_{22}(z)
\end{pmatrix}
\end{align*}
with $\Psi_{22}(z)$ a priori fixed and $\Psi_{11}(z)$, $\Psi_{12}(z)$ linearly parameterized. 

\section{Guarantees for the Infinite Horizon}
\label{sec:infinite}
While already many data-based methods exist which learn control theoretic system properties holding over a finite time horizon (cf.\ $L$-IQC in Def.~\ref{def:L_IQC}) \cite{Maupong2017,Montenbruck2016,Oomen2014,Rojas2012,Romer2019a,Romer2019c}, there have only been very few results on the infinite horizon (cf.\ Def.~\ref{def:iqc}) given finite data. While sharp results on the infinite horizon without an explicit model of the system seem difficult, it is already sufficient to receive an upper bound on the difference between infinite and finite horizon to guarantee a (possibly conservative) system property over the infinite horizon. Informally speaking, we investigate how much can go wrong by only considering the finite horizon in Def.~\ref{def:L_IQC}. We start with quite general results for IQCs satisfied by single-input single-output (SISO) systems and then explain what this implies for specific IQCs and dissipation inequalities. We will end this section with a short discussion on MIMO systems. 

We start with the following lemma that establishes a transformation, which will later be used in the main result of this section. 
\begin{lem}
\label{lem:stablePsi}
Let $G$ be a stable, discrete-time LTI system and let $\Psi$ be a causal and stable LTI filter. If there exists a stable and causal left inverse $\Psi_{11}^{-1}$ such that $\| \Psi_{12} G \Psi_{11}^{-1} \|_\infty < 1$, then $\widetilde{G}: r_1 \mapsto r_2$ as depicted in Fig.~\ref{fig.transfer} is causal and stable. 
\end{lem}
\begin{figure}[h]
\begin{center}
\begin{tikzpicture}
\draw[thick, fill=white!10] (0,0) rectangle (1.5,1);
\draw[thick] (0,-1.5) rectangle (1.5,-.5);
\draw[thick, fill=white!10] (3.5,0) rectangle (5.0,1);
\draw[thick] (3.5,-1.5) rectangle (5.0,-.5);
\draw[thick,->] (1.5,0.5) -- (3.5,0.5);
\draw[thick,->] (5,0.5) -- (5.65,0.5);
\draw[thick,->] (5.85,0.5) -- (6.5,0.5);
\node[anchor=south east] at (-1.25,0.5) {\small $r_1$};
\node[anchor=south east] at (6.75,0.5) {\small $r_2$};
\draw[thick,-] (0,-1) -- (-.75,-1);
\draw[thick] (-.75,0.5) circle(0.1);
\node[anchor=north east] at (-.75,0.5) {\small $-$};
\draw[thick] (5.75,0.5) circle(0.1);
\node[anchor=north east] at (6.25,0.5) {\small $+$};
\draw[thick,-] (2.25,0.5)--(2.25,-1);
\draw[thick,-] (2.75,0.5)--(2.75,-1);
\draw[thick,->] (-1.5,.5) -- (-.85,0.5);
\draw[thick,->] (-.75,-1) -- (-.75,0.4);
\draw[thick,->] (5.75,-1) -- (5.75,0.4);
\draw[thick,->] (-0.65,0.5) -- (0,0.5);
\draw[thick,->] (2.25,-1) -- (1.5,-1);
\draw[thick,->] (2.75,-1) -- (3.5,-1);
\draw[thick,-] (5,-1) -- (5.75,-1);
\node at (4.25,0.5) {\small $\Psi_{21}$};
\node at (4.25,-1) {\small $\Psi_{22}G$};
\node at (0.75,0.5) {\small $\Psi_{11}^{-1}$};
\node at (0.75,-1) {\small $\Psi_{12}G$};
\end{tikzpicture}
\end{center}
\caption{Transfer function $\widetilde{G}: r_1 \mapsto r_2$.}
\label{fig.transfer}
\end{figure}
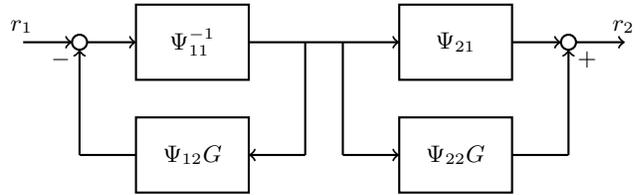
\begin{pf}
By assumption, $\Psi_{21} + \Psi_{22} G$ is causal and stable. With the small-gain theorem, we know that $\left( \Psi_{11} + \Psi_{12} G \right)^{-1}$ is causal and stable if $\| \Psi_{12} G \Psi_{11}^{-1} \|_\infty < 1$. Note that the small-gain argument also implies well-posedness of the interconnection in Fig.~\ref{fig.transfer}. As a series of causal and stable LTI systems, $\widetilde{G}: r_1 \mapsto r_2$ as depicted in Fig.~\ref{fig.transfer} is causal and stable. \qed
\end{pf}
This lemma leads us to the main result of this section for SISO LTI systems.
\subsection{Infinite Horizon Properties of SISO Systems}
For SISO LTI systems, where $m=p=n_{r_1}=n_{r_2}=1$, we can bound the remainder term from an $L$-IQC to an IQC over the infinite horizon using \cite{Boettcher2000,Tu18} via the following result.
\begin{thm}
\label{thm:finite}
Let $G$ be a stable, discrete-time SISO LTI system, which satisfies the $L$-IQC given by $P(z) = \Psi^{\sim}(z) M \Psi(z)$ with
\begin{align*}
M = \begin{pmatrix} \gamma^2 & \phantom{-}0 \\ 0 & -1 \end{pmatrix}
\end{align*}
and let $\Psi$ admit a stable and causal inverse of $\Psi_{11}$ such that $\| \Psi_{12} G \Psi_{11}^{-1} \| < 1$. Then, $G$ satisfies the (infinite horizon) IQC given by $P(z) = \Psi^\sim (z) M_{\inf} \Psi(z)$ with
\begin{align*}
M_{\inf} = \begin{pmatrix} (\gamma + \epsilon)^2 & \phantom{-}0 \\ 0 & -1 \end{pmatrix}
\end{align*}
and $\epsilon = \mathcal{O} \left( \frac{1}{L^2} \right)$.
\end{thm}
\begin{pf}
We introduce
\begin{align*}
\begin{pmatrix} r_1 \\ r_2 \end{pmatrix}
= 
\begin{pmatrix} \Psi_{11} & \Psi_{12} \\ \Psi_{21} & \Psi_{22} \end{pmatrix}
\begin{pmatrix} u \\ y \end{pmatrix}.
\end{align*} 
With $u = (\Psi_{11} + \Psi_{12} G )^{-1} r_1$ and $r_2 = (\Psi_{21} + \Psi_{22}G) u$, $\widetilde{G}: r_1 \mapsto r_2$ can be depicted as given in Fig.~\ref{fig.transfer}. 
Lem.~\ref{lem:stablePsi} shows that the transformed system $\widetilde{G}$ is causal and stable, which reduces the problem to finding the difference of the $\L2$-gain of $\widetilde{G}$ over the horizon $L$ and the $\L2$-gain of $\widetilde{G}$ over the infinite time horizon. Since we showed that $\widetilde{G}$ is causal and stable, we can apply \cite[Thm 4.1]{Boettcher2000} to prove that the $\mathcal{L}_2$-gain over the horizon $L$ of $\widetilde{G}$ approaches the $\mathcal{L}_2$-gain over the infinite horizon for increasing $L$ with $\epsilon = \mathcal{O} \left( \frac{1}{L^2} \right)$. \hfill \qed 
\end{pf}
The general idea of Thm.~\ref{thm:finite} is hence to translate the problem of a desired IQC for a system $G$ to the question of the $\mathcal{L}_2$-gain of a transformed (but stable and causal) system $\widetilde{G}$ and then to apply the results in \cite{Boettcher2000} for the $\L2$-gain. 
A specific expression for $\epsilon$ and a lower bound on $L$ to retrieve a specific $\epsilon$ can be found in \cite{Tu18} for all $L \geq 3$. Thm.~\ref{thm:finite} implies that the difference between considering a finite versus an infinite horizon vanishes at least quadratically with increasing horizon $L$. 

The results above apply, for example, 
to the most common dissipativity properties. 
Choose, for example, $\Psi_c$ as in \eqref{eq:conic_filter}
to retrieve general conic relations. 
This yields for the radius of the conic relation $\gamma_{\text{infinite}} \leq \gamma_{\text{finite}} + \epsilon$. Furthermore, since $\epsilon$ depends on the norm of $\widetilde{G}$, finding the minimal cone containing the input-output operator typically reduces also the remainder term $\epsilon$.
Other examples include output strict passivity where the $\L2$-gain estimation over the infinite time horizon of the transformed system via
\begin{align*}
\Psi_{\rho_{\text{o}}} {=} \begin{pmatrix} 1 & 0 \\ (\gamma + \epsilon)  & -\frac{1}{2 (\gamma+\epsilon)} \end{pmatrix}
\end{align*}
yields the output strict passivity property over the infinite time horizon.  
Note that $\Psi_{\text{c}}$ as well as $\Psi_{\rho_\text{o}}$ immediately satisfy the conditions of Thm.~\ref{thm:finite}, i.e. admit a stable and causal inverse of $\Psi_{11}$ and $\| \Psi_{12} G \Psi_{11}^{-1} \|_\infty = 0$.

\begin{rem}
Thm.~\ref{thm:finite} combined with the results from \cite{Tu18} suggest roughly that smaller $\| \widetilde{G} \|_{\infty}$ imply smaller $\epsilon$ for a given data length $L$. To decrease $\epsilon$ or in order to use shorter input-output trajectories, a very interesting filter $\Psi$ is 
\begin{align*}
\Psi = \begin{pmatrix}
I & 0 \\ 0 & \mathcal{S}^{\rho-}
\end{pmatrix},
\end{align*}
where $\mathcal{S}^{\rho-}$ is defined by $w^\rho = \mathcal{S}^{\rho-} w$ as $w^\rho_k = \rho^k w_k$, $\rho<1$. 
For more information, the reader is referred to \cite{Hu2016}, where this filter has been used to show exponential stability. Applying these exponential filters to input-output trajectories can decrease $\epsilon$ and can extend the results of this section to unstable systems, where the numerical calculations otherwise become difficult. Special attention, however, must be paid to how this is influenced by measurement noise.
\end{rem}
The restricting condition for the application of Thm.~\ref{thm:finite} to general IQCs is that we assume $n_r = n_s = 1$ although many IQC descriptions include tall filters $\Psi$. Generally speaking, $n_r, n_s > 1$ brings us to the MIMO problematic as discussed in the next subsection. 

\subsection{Infinite Horizon Properties of MIMO Systems}
Since the result from \cite{Tu18} and hence the results above strongly rely on the Toeplitz structure of an input-output description of a discrete time SISO LTI system and the corresponding results on the norms of Toeplitz matrices \cite{Boettcher2000} they are not transferable to MIMO systems. An obvious, but quite conservative, approach is to bound the $\L2$-gain over the infinite time horizon from each of the inputs to each of the outputs and take the matrix norm as an upper bound for the $\L2$-gain over the infinite time horizon of the MIMO system. In most cases, however, this yields a quite conservative bound for the $\L2$-gain of the MIMO system. 

Very generally, we will retrieve the exact operator norm with Thm.~\ref{thm:data_driven_IQC} in the limit $L \rightarrow \infty$ also for MIMO systems (cf. Appendix). However, we are more particularly interested how fast we approach the true operator gain (and then, via transformations as introduced before, any other system property). For general MIMO systems, this remains an open question. We have a very special solution for quadratic MIMO finite impulse response (FIR) models, which goes back to results on norms of finite sections of perturbed Toeplitz band matrices \cite{Rogozhin2005}, where the authors analyze how singular values of the finite sections Toeplitz matrices approximate the singular values of infinite Toeplitz matrices.

\begin{prop}
\label{thm:mimo_fir}
Given a square FIR MIMO system $G$,
\begin{align*}
G(z) = \sum_{k=0}^{l} g_k z^{-k}, \quad g_k \in \mathbb{R}^{m\times m}, l \in \mathbb{N},
\end{align*}
with $\text{det} \; G(z) \neq 0$ for all $\{z \in \mathbb{C}: |z| = 1\}$. 
Then, for all $L \geq 20 l$, $L$-dissipativity with
\begin{align*}
P = \begin{pmatrix} \gamma^2 I_m & \phantom{-}0 \\ 0 & -I_m \end{pmatrix} 
\end{align*}
implies dissipativity over the infinite horizon for 
\begin{align*}
P = \begin{pmatrix} (\gamma + \epsilon)^2 I_m & \phantom{-}0 \\ 0 & -I_m \end{pmatrix} 
\end{align*}
with $\epsilon \leq \frac{20 l}{L} \|G\|_\infty$. 
\end{prop}
\begin{pf}
This result is based on Lemma 4.3 in \cite{Rogozhin2005}. A more detailed explanation can be found in the Appendix.
\end{pf}
Together with Lem.~\ref{lem:stablePsi} from before, this can again be applied to other system properties. However, the transformed system needs to be again FIR, which holds for dissipativity properties but is generally not true in the case of general IQCs. 
While the above result is quite conservative it exemplarily shows that even in the MIMO case at least some qualitative results can be obtained.
\begin{exmp}
We determine the $\mathcal{L}_2$-gain and the smallest cone containing the input-output behavior of a randomly generated $2 \times 2$ MIMO system with system order $n=5$ (i.e. Matlab function \textit{drss} with \textit{rng(0)}) over different horizons $L$. 
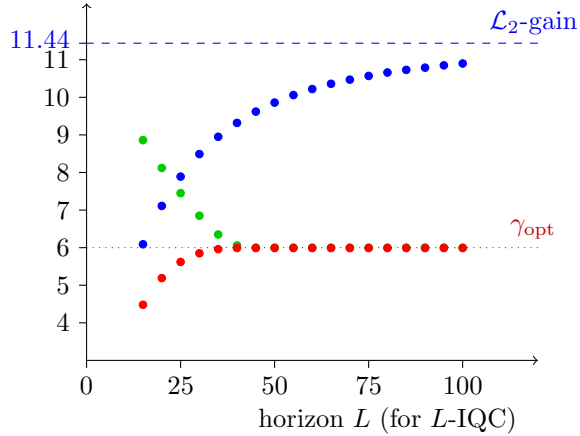
\begin{figure}[t]
\begin{tikzpicture}
 \newcommand*{\ax}{0.5}
 \newcommand*{\ay}{.5} 
 \newcommand*{\dx}{0}
 \draw[->] (0,1.5) -- (6,1.5) node[anchor=west] {};
 \draw[->] (0,1.5) -- (0,6.25) node[anchor=south,xshift=48,yshift=-15]{};  
\draw (0,\ay*4) -- (-.1,\ay*4) node[anchor=east,yshift=0] {$4$};
\draw (0,\ay*5) -- (-.1,\ay*5) node[anchor=east,yshift=0] {$5$};
\draw (0,\ay*6) -- (-.1,\ay*6) node[anchor=east,yshift=0] {$6$};
\draw (0,\ay*7) -- (-.1,\ay*7) node[anchor=east,yshift=0] {$7$};
\draw (0,\ay*8) -- (-.1,\ay*8) node[anchor=east,yshift=0] {$8$};
\draw (0,\ay*9) -- (-.1,\ay*9) node[anchor=east,yshift=0] {$9$};
\draw (0,\ay*10) -- (-.1,\ay*10) node[anchor=east,yshift=0] {$10$};
\draw (0,\ay*11) -- (-.1,\ay*11) node[anchor=east,yshift=0] {$11$};

\draw (\dx,1.5) -- (\dx,1.4) node[anchor=north] {$0$};
\draw (\ax*5+\dx,1.5) -- (\ax*5+\dx,1.4) node[anchor=north] {$50$};
\draw (\ax*10+\dx,1.5) -- (\ax*10+\dx,1.4) node[anchor=north] {$100$};
\draw (\ax*2.5+\dx,1.5) -- (\ax*2.5+\dx,1.4) node[anchor=north] {$25$};
\draw (\ax*7.5+\dx,1.5) -- (\ax*7.5+\dx,1.4) node[anchor=north] {$75$};
\node[anchor=north] at (\ax*8+\dx,1) {horizon $L$ (for $L$-IQC)};
\filldraw[blue] (\ax*1.5+\dx,\ay*6.09) circle (.05);
\filldraw[green!80!black] (\ax*1.5+\dx,\ay*8.86) circle (.05);
\filldraw[red] (\ax*1.5+\dx,\ay*4.48) circle (.05);
\filldraw[blue] (\ax*2+\dx,\ay*7.11) circle (.05);
\filldraw[green!80!black] (\ax*2+\dx,\ay*8.12) circle (.05);
\filldraw[red] (\ax*2+\dx,\ay*5.19) circle (.05);
\filldraw[blue] (\ax*2.5+\dx,\ay*7.89) circle (.05);
\filldraw[green!80!black] (\ax*2.5+\dx,\ay*7.45) circle (.05);
\filldraw[red] (\ax*2.5+\dx,\ay*5.62) circle (.05);
\filldraw[blue] (\ax*3+\dx,\ay*8.49) circle (.05);
\filldraw[green!80!black] (\ax*3+\dx,\ay*6.85) circle (.05);
\filldraw[red] (\ax*3+\dx,\ay*5.85) circle (.05);
\filldraw[blue] (\ax*3.5+\dx,\ay*8.95) circle (.05);
\filldraw[green!80!black] (\ax*3.5+\dx,\ay*6.35) circle (.05);
\filldraw[red] (\ax*3.5+\dx,\ay*5.96) circle (.05);
\filldraw[blue] (\ax*4+\dx,\ay*9.32) circle (.05);
\filldraw[green!80!black] (\ax*4+\dx,\ay*6.06) circle (.05);
\filldraw[red] (\ax*4+\dx,\ay*5.99) circle (.05);
\filldraw[blue] (\ax*4.5+\dx,\ay*9.62) circle (.05);
\filldraw[green!80!black] (\ax*4.5+\dx,\ay*6) circle (.05);
\filldraw[red] (\ax*4.5+\dx,\ay*5.99) circle (.05);
\filldraw[blue] (\ax*5+\dx,\ay*9.86) circle (.05);
\filldraw[green!80!black] (\ax*5+\dx,\ay*6) circle (.05);
\filldraw[red] (\ax*5+\dx,\ay*5.99) circle (.05);
\filldraw[blue] (\ax*5.5+\dx,\ay*10.06) circle (.05);
\filldraw[green!80!black] (\ax*5.5+\dx,\ay*6) circle (.05);
\filldraw[red] (\ax*5.5+\dx,\ay*5.99) circle (.05);
\filldraw[blue] (\ax*6+\dx,\ay*10.22) circle (.05);
\filldraw[green!80!black] (\ax*6+\dx,\ay*6) circle (.05);
\filldraw[red] (\ax*6+\dx,\ay*5.99) circle (.05);
\filldraw[blue] (\ax*6.5+\dx,\ay*10.36) circle (.05);
\filldraw[green!80!black] (\ax*6.5+\dx,\ay*6) circle (.05);
\filldraw[red] (\ax*6.5+\dx,\ay*5.99) circle (.05);
\filldraw[blue] (\ax*7+\dx,\ay*10.47) circle (.05);
\filldraw[green!80!black] (\ax*7+\dx,\ay*6) circle (.05);
\filldraw[red] (\ax*7+\dx,\ay*5.99) circle (.05);
\filldraw[blue] (\ax*7.5+\dx,\ay*10.57) circle (.05);
\filldraw[green!80!black] (\ax*7.5+\dx,\ay*6) circle (.05);
\filldraw[red] (\ax*7.5+\dx,\ay*5.99) circle (.05);
\filldraw[blue] (\ax*8+\dx,\ay*10.66) circle (.05);
\filldraw[green!80!black] (\ax*8+\dx,\ay*6) circle (.05);
\filldraw[red] (\ax*8+\dx,\ay*5.99) circle (.05);
\filldraw[blue] (\ax*8.5+\dx,\ay*10.73) circle (.05);
\filldraw[green!80!black] (\ax*8.5+\dx,\ay*6) circle (.05);
\filldraw[red] (\ax*8.5+\dx,\ay*5.99) circle (.05);
\filldraw[blue] (\ax*9+\dx,\ay*10.79) circle (.05);
\filldraw[green!80!black] (\ax*9+\dx,\ay*6) circle (.05);
\filldraw[red] (\ax*9+\dx,\ay*5.99) circle (.05);
\filldraw[blue] (\ax*9.5+\dx,\ay*10.85) circle (.05);
\filldraw[green!80!black] (\ax*9.5+\dx,\ay*6) circle (.05);
\filldraw[red] (\ax*9.5+\dx,\ay*5.99) circle (.05);
\filldraw[blue] (\ax*10+\dx,\ay*10.90) circle (.05);
\filldraw[green!80!black] (\ax*10+\dx,\ay*6) circle (.05);
\filldraw[red] (\ax*10+\dx,\ay*5.99) circle (.05);
\draw (0,\ay*11.44) -- (-.1,\ay*11.44) node[anchor=east,yshift=1.5, color=blue!95!black] {$11.44$};
\draw[blue!95!black, dashed] (6,\ay*11.44) -- (-.1,\ay*11.44) node[pos=.01,anchor=south] {$\L2$-gain};
\draw[red!75!black, dotted] (6,\ay*6) -- (-.1,\ay*6) node[pos=.01,anchor=south] {$\gamma_{\text{opt}}$};
\end{tikzpicture}
\caption{Determining the $\L2$-gain and the tightest cone containing the input-output behavior of an $2 \times 2$ MIMO system over different horizons $L$. Here, \textcolor{blue}{$\bullet$}
denotes the $\L2$-gain estimate, \textcolor{red}{$\bullet$} denotes the $\gamma$ estimate resulting from the corresponding $C$ estimate, and \textcolor{green}{$\bullet$} denotes the true $\gamma$ with respective to the same $C$ estimate.}
\label{fig:ex_sp}
\end{figure}
The results in Fig.~\ref{fig:ex_sp} illustrate different phenomena. Even though the theoretical results in Thm.~\ref{thm:finite} hold only for SISO systems, it seems that similar behavior can be observed also for MIMO systems. Furthermore, by allowing for a transformation with a center matrix $C$, we can greatly reduce the conservatism of the $\mathcal{L}_2$-gain. Moreover, as suggested in our analysis, due to the smaller $\mathcal{L}_2$-gain of the transformed system, the radius $\gamma_{\mathrm{opt}}$ converges with increasing $L$ significantly faster towards the corresponding infinite horizon value. During the first steps, as the optimal $C$ parameter also deviates from the optimal $C$ due to the short data horizon, the minimal radius for the non-optimal $C$ is larger than the minimal radius for the optimal $C$.
\end{exmp}

\section{High Dimensional Numerical Example}
\label{sec:example}
Throughout the last sections, we provided small examples to illustrate the presented results and effects of different parameters. In the following, we focus on a rather high dimensional system with connection to a real application. We consider a model of a building (the Los Angeles University Hospital), which has been listed as a benchmark model reduction problem, e.g., in \cite{Chahlaoui2002,Tran2017} when a full mathematical model of the system is known. The model can be found, for example in \cite{Tran2017} and references therein\footnote{The authors of \cite{Tran2017} made their MATLAB files available on \href{http://verivital.com/hyst/pass-order-reduction/}{http://verivital.com/hyst/pass-order-reduction/}.}. The building has eight floors each with three degrees of freedom. The model is hence of dimension $n=48$ and we overestimate the system order by $\nu = 50$. We simulate the model with a sampling time of $\delta t=0.1$. The SDPs are solved with the Multi-Parametric Toolbox 3.0 \cite{Herceg2013} together with YALMIP \cite{Loefberg2004}. The results presented below illustrate that the presented method is simple to apply, can outperform model identification via standard tools and consecutive analysis of the identified model, and it provides good results in determining IQCs even for high dimensional systems and in the presence of measurement noise. 

\subsection{Dissipativity Properties - a comparison to system identification methods}
We measure $K=3$ trajectories of length $N=2400$ with the output being subject to uniform multiplicative measurement noise of the form $\tilde{y_k} = (1+ \varepsilon_k) y_k$ with $\varepsilon_k \in [-\bar{\varepsilon}, \bar{\varepsilon}]$ at different noise levels $\bar{\varepsilon} > 0$ representing the signal to noise ration (SNR).
The true system has an input-feedforward passivity parameter of $\rho_{\text{i}}=-0.001$
and, as also stated in \cite{Chahlaoui2002}, an operator gain of $\gamma = 0.0052$. The following table provides the estimated operator gain and passivity index via simple bisection method together with 
Algorithm 1 for $L=1050$. It can be seen that even for high levels of noise, the estimate of the operator gain as well as the passivity parameter are consistently close to the true values.

\begin{center}
\scriptsize
\begin{tabular}{|l||c|c|c|c|c|}
\hline
$\bar{\varepsilon}$ & 0 & 1 \% & 10 \% & 25 \% & 50 \% \\ \hline \hline
$\gamma$ ($\cdot 10^{-3}$) & 5.2 & 5.2 & 5.1 & 5.1 & 5.0 \\ \hline
$\rho_i$ ($\cdot 10^{-3}$) & -1.0 & -1.0 & -0.9 & -0.9 & -0.8 \\ \hline
\end{tabular}
\end{center}

We now choose the same input-output trajectory for the case $\bar{\varepsilon} = 0.25$ and apply standard system identification tools, i.e. \textsc{Matlab} functions \textit{ssest} (estimates state-space model by initializing the parameter via a subspace approach or an iterative rational function estimation approach and then refines the parameter values using the prediction error minimization approach), \textit{ssregest} (estimates state-space model by reduction of a regularized ARX model), and \textit{n4sid} (estimates state-space model using a subspace method). Each of the approaches are initialized with three different assumptions on the model order: 10 (underestimated), 41 (suggested by the respective \textsc{Matlab} function), 70 (overestimated).  After the model identification, we then determine the gain and the input feedforward passivity index with \textit{norm($\cdot$,inf), getPassiveIndex($\cdot$,'input')}, respectively. The result is summarized in the table below.

 \begin{center}
\scriptsize
{\renewcommand{\arraystretch}{1.9}%
 \begin{tabular}{|l||lll|}
 \hline
\pbox{2.6cm}{assumed \\ system order:} & \pbox{3cm}{10}  & \pbox{3cm}{41} & \pbox{3cm}{70} 
\\[1ex]
 \hline
\hline
\pbox{1.9cm}{\textit{ssest} ($\cdot 10^{-3}$)} 
& \pbox{2cm}{$\gamma = 5.3$ \\ $\rho_{\text{i}} = -0.2$ } 
& \pbox{2cm}{$\gamma = 5.9$ \\ $\rho_{\text{i}} = -0.3$ } 
& \pbox{2cm}{$\gamma = 5.3$ \\ $\rho_{\text{i}} = -\infty$ } 
\\[2ex]
\pbox{2.1cm}{\textit{ssregest} ($\cdot 10^{-3}$)}
& \pbox{2cm}{$\gamma = 4.4$ \\ $\rho_{\text{i}} = -0.8$ } 
& \pbox{2cm}{$\gamma = 5.1$ \\ $\rho_{\text{i}} = -1.1$ }
& \pbox{2cm}{$\gamma = 5.3$ \\ $\rho_{\text{i}} = -1.0$ }
\\[2ex]
\pbox{2.1cm}{\textit{n4sid} ($\cdot 10^{-3}$)} 
& \pbox{2cm}{$\gamma = 5.4$ \\ $\rho_{\text{i}} = -0.8$ }
& \pbox{2cm}{$\gamma = 5.3$ \\ $\rho_{\text{i}} = -1.0$ }
& \pbox{2cm}{$\gamma = 7.6$ \\ $\rho_{\text{i}} = -\infty$ }
\\[1ex]
 \hline
 \end{tabular}}
 \end{center}
We can see that standard system identification tools from one noise-corrupted input-output trajectory produced quite variable results which are also highly dependent on the assumed system order. 
Please also note that for a system order of $70$ the system identification function \textit{ssest} required over $1$ hour on an Intel i7, while the computational expenses for the simple bisection method together with Algorithm~1 are below one minute.

\subsection{Determining an optimal IQC}
From expert knowledge and insights to the mechanics of the building, one might have the suspicion that part of the system dynamics of the high dimensional system approximately behaves like a second order low-pass filter. With the presented method, one can apply a data-driven low-order approximation approach with - most importantly - a guaranteed bound on the approximation error $\gamma$. We choose the IQC 
\begin{align*}
M = \begin{pmatrix} \gamma^2 I_{m} & 0 \\ 0 & -I_{p} \end{pmatrix} \quad
\Psi(z) &= \begin{pmatrix} I_{m} & 0 \\ -\Psi_{21}(z) & I_{p} \end{pmatrix}
\end{align*}
with $\Psi_{21}(z) = \sum_{k=1}^2 c^{(21)}_k B^{(21)}_k(z)$, where we 
choose two second order low-pass filters as basis functions.
As a result, we hence find with $N=1210$ that over the horizon $L-\nu = 500$ ($L=550$), the best approximation $G_{\text{lo}}$ with the chosen basis functions $B^{(21)}_k(z)$, $k=1,2$ is 
\begin{align*}
G_{\text{lo}} = \frac{2.67\cdot 10^{-4}(10z + 1)}{z^2 + 0.5 z + 0.1} + \frac{5.33\cdot 10^{-5}(z+1)}{z^2 - 1.2 z + 0.7}
\end{align*}
and the difference between $G_{\text{full}}$ and the lower order $G_{\text{lo}}$ over the horizon $L-\nu = 500$ is bounded by $\gamma = 0.0035$.
As reference value, we find with full model knowledge $\|G_{\text{full}} - G_{\text{lo}} \|_\infty = 0.0035$. Finding and verifying an IQC hence worked via simple computations. 

In the following we investigate how the choice of the horizon as well as $\nu$ influences the computed best approximation $G_{\text{lo}}$ as well as the radius $\gamma$. 
The table shows $\|G_{\text{full}} {-} G_{\text{lo}} \|_\infty$ for the computed best approximation $G_{\text{lo}}$ and the computed radius $\gamma$ for the corresponding $L-\nu$-IQC. 
We can see that shortening the horizon does not significantly deteriorate the computed $G_{\text{lo}}$, and even for this high-dimensional system $\gamma$ quickly approaches the corresponding value of the infinite horizon for increasing $L$. 

\begin{center}
\scriptsize
\begin{tabular}{|l||c|c|c|c|c|}
\hline
$L-\nu$ & 100 & 200 & 300 & 400 & 500 \\ \hline \hline
$\|G_{\text{full}} {-} G_{\text{lo}} \|_\infty$ {\tiny ($\cdot 10^{-3}$)} & 3.7 & 3.6 & 3.5 & 3.5 & 3.5 \\ \hline
$\gamma$ {\tiny ($\cdot 10^{-3}$)} & 2.9 & 3.3 & 3.4 & 3.4 & 3.5 \\ \hline
\end{tabular}
\end{center}

As implied by the theoretical results in Sec.~\ref{sec:main}, there is no change in the result for different $\nu \geq n$. 
Even for smaller $\nu$, we receive an upper bound on the true $L-\nu$-IQC, which might however be conservative (especially for $\nu \ll n$), as explained in Sec.~\ref{sec:main}.

\begin{center}
\scriptsize
\begin{tabular}{|l||c|c|c|c|c|} 
\hline
$\nu$ & 10 & 28 & 48 & 50 & 70\\ \hline \hline
$\|G_{\text{full}} {-} G_{\text{lo}} \|_\infty$ {\tiny ($\cdot 10^{-3}$)} & 5.2 & 4.3 & 3.5 & 3.5 & 3.5 \\ \hline
$\gamma$ {\tiny ($\cdot 10^{-3}$)} & 65.2 & 4.4 & 3.5 & 3.5 & 3.5\\ \hline
\end{tabular}
\end{center}

Finally, we add measurement noise and calculate the approximation error $\gamma$ considering the IQC with the above given parameters $c$ via Algorithm~1 and $K=3$. We can see that even for high dimensional systems, verifying an IQC works remarkably well also in the presence of significant measurement noise.
\begin{center}
\scriptsize
\begin{tabular}{|l||c|c|c|c|c|c|}
\hline
$\bar{\varepsilon}$ & 0 \% & 1 \% & 5 \% & 10 \% & 25 \% & 50 \% \\ \hline \hline
$\gamma$ {\tiny ($\cdot 10^{-3}$)} & 3.5 & 3.5 & 3.5 & 3.4 & 3.4 & 3.4 \\ \hline
\end{tabular}
\end{center}

\section{Conclusion and Outlook}
We introduced a simple approach to verify and find IQCs from only one input-output trajectory of an unknown LTI system. We also provided guarantees to find the tightest system property descriptions via a simple SDP, as well as introduced means to infer the system properties also over the infinite horizon. A high-dimensional application example showed the potential of the approach even for challenging applications as well as in the presence of noise. Generalizing these ideas to (slightly) nonlinear systems shall be the subject of future work.

\begin{ack}                               
The authors thank the German Research Foundation (DFG) for financial support of the project within the German Excellence Strategy - EXC 2075 - 390740016, along with the Max Planck Research School (IMPRS) for Intelligent Systems for their support.  
\end{ack}

\bibliographystyle{plain}        
\bibliography{../../Literature}     
                                
\appendix

\section{Appendix}
In the following, we briefly summarize some results from \cite{Boettcher2000} and \cite{Rogozhin2005} which can then be used to prove Prop.~\ref{thm:mimo_fir}.

Consider a sequence $\{a_n\}_{n= -\infty}^\infty$ of complex numbers and the corresponding Toeplitz matrix 
\begin{align}
A = \begin{pmatrix} a_0 & a_{-1} & a_{-2} & \dots \\
a_1 & a_{0} & a_{-1} & \dots \\
a_2 & a_{1} & a_{0} & \dots \\
\dots & \dots & \dots & \dots
\end{pmatrix}.
\label{eq:Toeplitz}
\end{align}
\begin{thm}[Toeplitz 1911]
\label{thm:toeplitz}
The matrix \eqref{eq:Toeplitz} defines a bounded operator on $l_2$ if and only if the numbers $\{a_n\}$ are Fourier coefficients of some function $a \in l_\infty$,
\begin{align*}
a_n = \frac{1}{2\pi} \int_0^{2\pi} a(e^{i \Theta})e^{-in\Theta} d \Theta, \quad n \in \mathbb{Z}, 
\end{align*}
where $i$ denotes the imaginary unit. In that case the norm of the operator given by \eqref{eq:Toeplitz} equals \begin{align*}
\|a\|_\infty := \text{ess sup}_{z \in \mathbb{C}: |z| = 1} |a(z)|.
\end{align*}
\end{thm}

Hence, the Toeplitz matrix $T(G)$ representing the convolution operator of an asymptotically stable SISO LTI system $G$ is a bounded operator with $\|T(G)\| = \|G\|_\infty$ representing its operator norm. Let us now consider quadratic MIMO systems $m=p$.

Let $a_k$, $k \in \mathbb{Z}$ be the $m \times m$ matrix $((a_{ij})_k)_{i,j = 1}^m$ formed by the Fourier coefficients of a function $a = (a_{ij})_{i,j=1}^m \in \mathcal{C}^{m \times m}$, where $C$ denotes all continuous functions defined on the unit circle $\{z \in \mathbb{C} : |z| = 1 \}$. The block Toeplitz operator $T(a): l^m_2 \rightarrow l^m_2$ is defined by the matrix representation
\begin{align*}
T(a) = (a_{i-j})_{i,j = 0}^\infty = \begin{pmatrix} a_0 & a_{-1} & a_{-2} & \dots \\
a_1 & a_{0} & a_{-1} & \dots \\
a_2 & a_{1} & a_{0} & \dots \\
\dots & \dots & \dots & \dots
\end{pmatrix},
\end{align*}
which can hence represent the convolution operator of MIMO systems.  
 
Considering the input-output behavior over a finite time horizon, the resulting input-output operator is a finite section of the Toeplitz operator $T(a)$. We define $S_L$ such that $S_L x = S_L(x_0, x_1, \dots, x_{L-1}, x_L, x_{L+1}, \dots) = (x_0, x_1, \dots, x_{L-1}, 0, 0, \dots)$. Then the finite sections $T_L(a)$ are defined by the truncated $mL \times mL$ matrices
\begin{align*}
T_L(a) := S_L T(a) S_L = (a_{i-j})_{i,j=0}^{L-1}, \quad L \in \mathbb{N}.
\end{align*}

From e.g. \cite{Rogozhin2005}, we then know that for all generating functions $a \in C^{m\times m}$, the Toeplitz operator $T(a)$ is a linear bounded operator on $l^m_2$ and $\|T(a)\| = \|a\|_{\infty}$. Hence, we'll retrieve the exact operator norm in the limit (for $L \rightarrow \infty$).

Furthermore, from \cite{Rogozhin2005} we know that a Toeplitz operator $T(a)$, $a \in C^{m\times m}$ is Fredholm if and only if $\text{det} \; a(z) \neq 0$ for all $\{z \in \mathbb{C}: |z| = 1\}$, which generally requires that $a(z)$ has no zeros on the unit circle.
This brings us to Lem.~4.3 in \cite{Rogozhin2005}.
\begin{lem}
Let $a$ be a trigonometric polynomial of the form
\begin{align*}
a(z) = \sum_{k=-l}^{l} a_k z^{-k}, \quad a_k \in \mathbb{C}^{m\times m}, l \in \mathbb{N},
\end{align*}
such that $\text{det} \; a(z) \neq 0$ for all $\{z \in \mathbb{C}: |z| = 1\}$. 
If $L \geq 20 l$ then 
\begin{align*}
\left( 1 - \frac{20l}{L} \right) \| a \|_\infty = \left( 1 - \frac{20l}{L} \right) \| T(a) \| \leq \| T_L(a) \|. 
\end{align*}
\end{lem}
Applying this result with $a_k \in \mathbb{R}^{m \times m}$, and $a_k = 0$ for all $k < 0$ brings us directly to Prop.~\ref{thm:mimo_fir} for MIMO FIR filters.

\end{document}